\documentclass[aps,prb,superscriptaddress,a4paper,showpacs, reprint, floatfix]
{revtex4-1}

\usepackage{graphicx}
\usepackage{amsmath}
\usepackage{amssymb}
\usepackage{chapterbib}
\usepackage[squaren]{SIunits}
\usepackage{hyperref}
\hypersetup{pdfauthor={Edwin Menzel et al.},pdftitle={Path Entanglement of Continuous-Variable Quantum Microwaves}}
  
\begin{document}
\title{Path Entanglement of Continuous-Variable Quantum Microwaves}

\author{E. P. \surname{Menzel}}
\email{Edwin.Menzel@wmi.badw-muenchen.de}
\affiliation{Walther-Mei{\ss}ner-Institut, Bayerische Akademie der
Wissenschaften, D-85748 Garching, Germany}
\affiliation{Physik-Department, Technische Universit{\"a}t M{\"u}nchen, D-85748
Garching, Germany}
\author{R. \surname{Di Candia}}
\affiliation{Departamento de Qu{\'i}mica F{\'i}sica, Universidad del Pa{\'i}s
Vasco UPV/EHU, 48080 Bilbao, Spain}
\author{F. Deppe}
\affiliation{Walther-Mei{\ss}ner-Institut, Bayerische Akademie der
Wissenschaften, D-85748 Garching, Germany}
\affiliation{Physik-Department, Technische Universit{\"a}t M{\"u}nchen, D-85748
Garching, Germany}
\author{P. Eder}
\affiliation{Walther-Mei{\ss}ner-Institut, Bayerische Akademie der
Wissenschaften, D-85748 Garching, Germany}
\affiliation{Physik-Department, Technische Universit{\"a}t M{\"u}nchen, D-85748
Garching, Germany}
\author{L. Zhong}
\affiliation{Walther-Mei{\ss}ner-Institut, Bayerische Akademie der
Wissenschaften, D-85748 Garching, Germany}
\affiliation{Physik-Department, Technische Universit{\"a}t M{\"u}nchen, D-85748
Garching, Germany}

\author{M. Ihmig}
\affiliation{Lehrstuhl f{\"u}r Integrierte Systeme, Technische Universit{\"a}t
M{\"u}nchen, D-80333 M{\"u}nchen, Germany}

\author{M. Haeberlein}
\affiliation{Walther-Mei{\ss}ner-Institut, Bayerische Akademie der
Wissenschaften, D-85748 Garching, Germany}
\affiliation{Physik-Department, Technische Universit{\"a}t M{\"u}nchen, D-85748
Garching, Germany}
\author{A. Baust}
\affiliation{Walther-Mei{\ss}ner-Institut, Bayerische Akademie der
Wissenschaften, D-85748 Garching, Germany}
\affiliation{Physik-Department, Technische Universit{\"a}t M{\"u}nchen, D-85748
Garching, Germany}
\author{E. Hoffmann}
\affiliation{Walther-Mei{\ss}ner-Institut, Bayerische Akademie der
Wissenschaften, D-85748 Garching, Germany}
\affiliation{Physik-Department, Technische Universit{\"a}t M{\"u}nchen, D-85748
Garching, Germany}

\author{D. Ballester}
\affiliation{Department of Physics and Astronomy, University College London,
London WC1E 6BT, United Kingdom}
\affiliation{Departamento de Qu{\'i}mica F{\'i}sica, Universidad del Pa{\'i}s
Vasco UPV/EHU, 48080 Bilbao, Spain}

\author{K. Inomata}
\affiliation{RIKEN Advanced Science Institute, Wako, Saitama 351-0198, Japan}
\author{T. Yamamoto}
\affiliation{NEC Smart Energy Research Laboratories, Tsukuba, Ibaraki, 305-8501,
Japan}
\affiliation{RIKEN Advanced Science Institute, Wako, Saitama 351-0198, Japan}
\author{Y. Nakamura}
\affiliation{Research Center for Advanced Science and Technology (RCAST), The
University of Tokyo, Komaba, Meguro-ku, Tokyo 153-8904, Japan}
\affiliation{RIKEN Advanced Science Institute, Wako, Saitama 351-0198, Japan}

\author{E. Solano}
\affiliation{Departamento de Qu{\'i}mica F{\'i}sica, Universidad del Pa{\'i}s
Vasco UPV/EHU, 48080 Bilbao, Spain}
\affiliation{IKERBASQUE, Basque Foundation for Science, 48011 Bilbao, Spain}
\author{A. Marx}
\affiliation{Walther-Mei{\ss}ner-Institut, Bayerische Akademie der
Wissenschaften, D-85748 Garching, Germany}
\author{R. Gross}
\affiliation{Walther-Mei{\ss}ner-Institut, Bayerische Akademie der
Wissenschaften, D-85748 Garching, Germany}
\affiliation{Physik-Department, Technische Universit{\"a}t M{\"u}nchen, D-85748
Garching, Germany}

\begin{abstract}
Path entanglement constitutes an essential resource in quantum information and
communication protocols. Here, we demonstrate frequency-degenerate entanglement
between continuous-variable quantum microwaves propagating along two spatially
separated paths. We combine a squeezed and a vacuum state using a microwave beam
splitter. Via correlation measurements, we detect and quantify the path
entanglement contained in the beam splitter output state. Our experiments open
the avenue to quantum teleportation, quantum communication, or quantum radar
with continuous variables at microwave frequencies.
\end{abstract}
\pacs{03.67.Bg, 03.65.Ud, 42.50.Dv, 85.25.-j}

\maketitle
\begin{cbunit}
Fascinatingly, quantum mechanics allows for a compound system to have a common
description while, at the same time, no individual states can be ascribed to its
subsystems~\cite{Horodecki:2009a}.
The presence of entanglement between spatially separated systems is a necessary
condition for what Einstein called ``spooky action at a
distance''~\cite{Born:1971a}: the contradiction between quantum mechanics and
local realism~\cite{Einstein:1935a,Horodecki:2009a}. Furthermore, entanglement
is at the heart of quantum communication and information processing
technologies, which promise significant performance gains over classical
protocols~\cite{Horodecki:2009a,Raimond:2001a,Braunstein:2005a}. Consequently,
entanglement has been extensively explored in atomic physics and quantum
optics~\cite{Ou:1992a,Raimond:2001a,Braunstein:2005a}. In these investigations,
optical frequencies were preferred over microwaves because the higher photon
energies facilitate practical applications. However, since the late 1990s,
microwave technology has evolved rapidly in  both industry and science. For one
thing, classical microwave fields have become an indispensable tool in mobile
communication. For another, a promising direction towards scalable quantum
information processing has appeared with the advent of superconducting microwave
quantum circuits~\cite{Schoelkopf:2008a,Clarke:2008a,Mariantoni:2011a}. Despite
some decoherence issues, these systems provide unprecedented light-matter
coupling strengths due to their large effective dipole moments and field
enhancement effects~\cite{Wallraff:2004a,Niemczyk:2010a}. 
As a consequence,
standing-wave fields in transmission line resonators were shown to act as a
short-range quantum bus between superconducting
qubits~\cite{Majer:2007a,Sillanpaa:2007a} and various gates were
implemented~\cite{Yamamoto:2003a,Sillanpaa:2007a,Majer:2007a,Plantenberg:2007a,
Ansmann:2009a}. For microwave quantum communication, however, propagating fields
are required. As a first step in this direction, early experiments demonstrated
tomography of weak thermal states~\cite{Mariantoni:2010a}, coherent
states~\cite{Menzel:2010a}, and single photons~\cite{Bozyigit:2011a}. Next,
continuous-variable states generated by Josephson parametric devices were
reconstructed~\cite{Mallet:2011a}. Very recently, such devices have permitted to
investigate two-mode squeezing~\cite{Eichler:2011b,Wilson:2011a,Flurin:2012a}.
An important aspect of these experiments is the understanding they provide
regarding entanglement. In order to be a resource in quantum communication
protocols, it must occur between spatially separated
subsystems~\cite{Horodecki:2009a}. Furthermore, a strict proof of entanglement
requires the entangler and the detector to be based on independent experimental
techniques.
\begin{figure*}
\centering
\includegraphics[width=\textwidth]{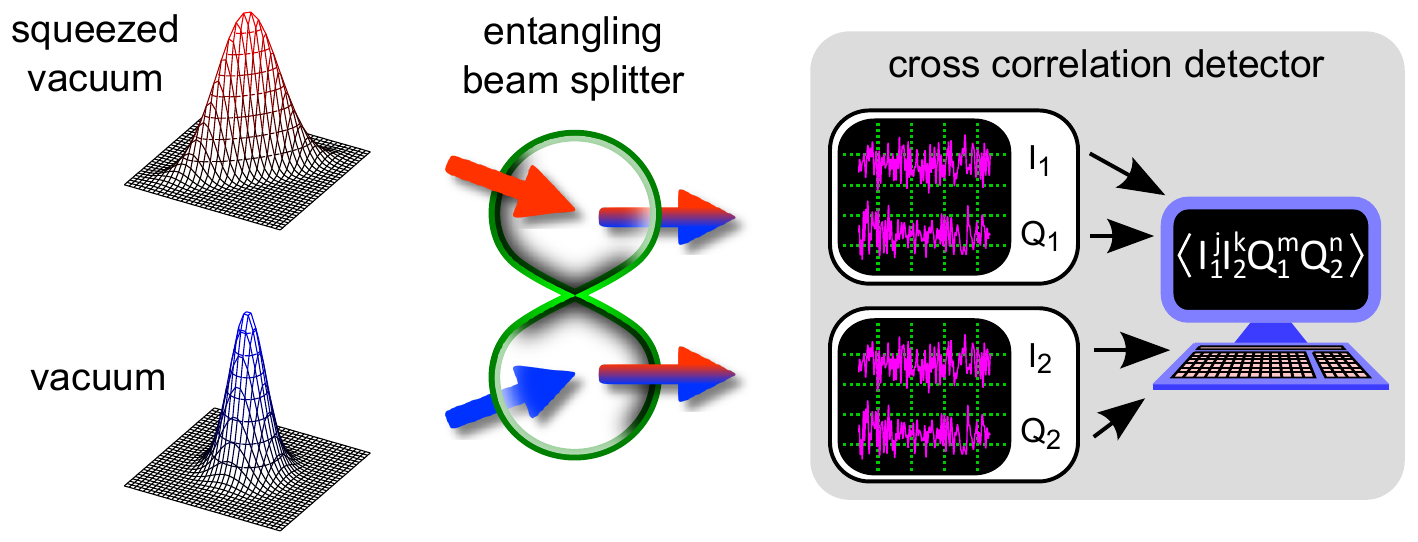}
\caption{\label{Path_Ent_Fig1}{Layout of the experiment.} The microwave beam
splitter acts as entangling device (green eight-shaped structure). The blue-and-
red arrows denote the path-entangled state. In the cross correlation detector,
the oscilloscope symbols denote the noisy amplification, down-conversion, and
digitizing of the in-phase ($I_{1,2}$) and quadrature ($Q_{1,2}$) components of
the output signals. The computer symbolizes the final numerical data processing,
partly done by an FPGA logic.}
\end{figure*}
In this work, we make a significant step beyond previous efforts and demonstrate
path entanglement in the microwave regime, respecting both criteria mentioned
above. Our experiments follow the spirit of the quantum-optical
realization~\cite{Ou:1992a} of the original Einstein-Podolsky-Rosen (EPR)
paradox~\cite{Einstein:1935a}. 
As shown in Fig.~\ref{Path_Ent_Fig1}, we combine
a vacuum and a squeezed vacuum state in a hybrid ring microwave beam
splitter~\cite{Hoffmann:2010a} acting as an entangling device. Its two output
ports hold a continuous-variable state which is frequency-degenerate and
entangled with respect to the two propagation paths. Along these paths, the
entanglement can be conveniently distributed to two parties requiring it for any
suitable quantum communication protocol. In our experiments, we first
reconstruct the squeezed \textit{input} state by means of dual-path
tomography~\cite{Menzel:2010a},
 which assumes knowledge of the beam splitter relations. Next, we reconstruct
 the moments of the \textit{output} state after the beam splitter by treating the
 latter as a black box and calibrating against a known
 state~\cite{Eichler:2011a}. In this reference-state
 method (see Supplementary), we only assume that independent vacuum states are
 produced in each output path when vacuum is incident at both input ports. From
 the moments reconstructed in this way, we build a witness matrix which proves
 the existence of path entanglement independently of the detailed nature of our
 output state~\cite{Shchukin:2005a}. Since in practice the data shows that our
 states are Gaussian, we finally quantify the degree of entanglement by means of
 the negativity~\cite{Adesso:2005a}. The result of this analysis agrees with
 what we expect for our squeezed input state. We note here that for bipartite
 single-mode Gaussian entanglement, as it is relevant in our case, entanglement
 implies nonlocality~\cite{Braunstein:2005a,Horodecki:2009a}. All in all, our
 results show that we have realized the main building block for microwave
 quantum teleportation and communication protocols.

The generation of the input states for the beam splitter is straightforward. The
vacuum is realized with a commercial 50\,\ohm-termination at 40\,mK acting as a
broadband black-body emitter~\cite{Mariantoni:2010a}. The squeezed state is
produced using a particular superconducting circuit, the flux-driven Josephson
parametric amplifier (JPA)~\cite{Yamamoto:2008a}. In this device two Josephson
junctions form a nonlinearity which can be modulated (``pumped'') at gigahertz
frequencies to achieve a parametric effect. The JPA box is stabilized to
$50$\,mK. A thermal state emitted by an  attenuator, whose temperature can be
varied from $50\,{-}\,800$\,mK, can be fed into the JPA. Our cross correlation
detector is based on the insight that for microwave signals off-the-shelf high-
gain low-noise linear amplifiers are available rather than efficient single
photon counters. We connect one amplification path to each output port of the
beam splitter. At room temperature we record the in-phase and quadrature
components, $I_{1,2}$ and $Q_{1,2}$ of the amplified signals. The averaged
moments $\langle I_1^jI_2^kQ_1^mQ_2^n\rangle$ are computed for $j\,{+}\,k\,{+}
\,m\,{+}\,n\,{\le}\,4$ and $j,k,m,n\,{\in}\,\mathbb{N}_0$ in real time using a
field programmable gate array (FPGA) logic. Further details can be found in the
Supplementary.

\begin{figure*}[t]
\centering
\includegraphics[width=\textwidth]{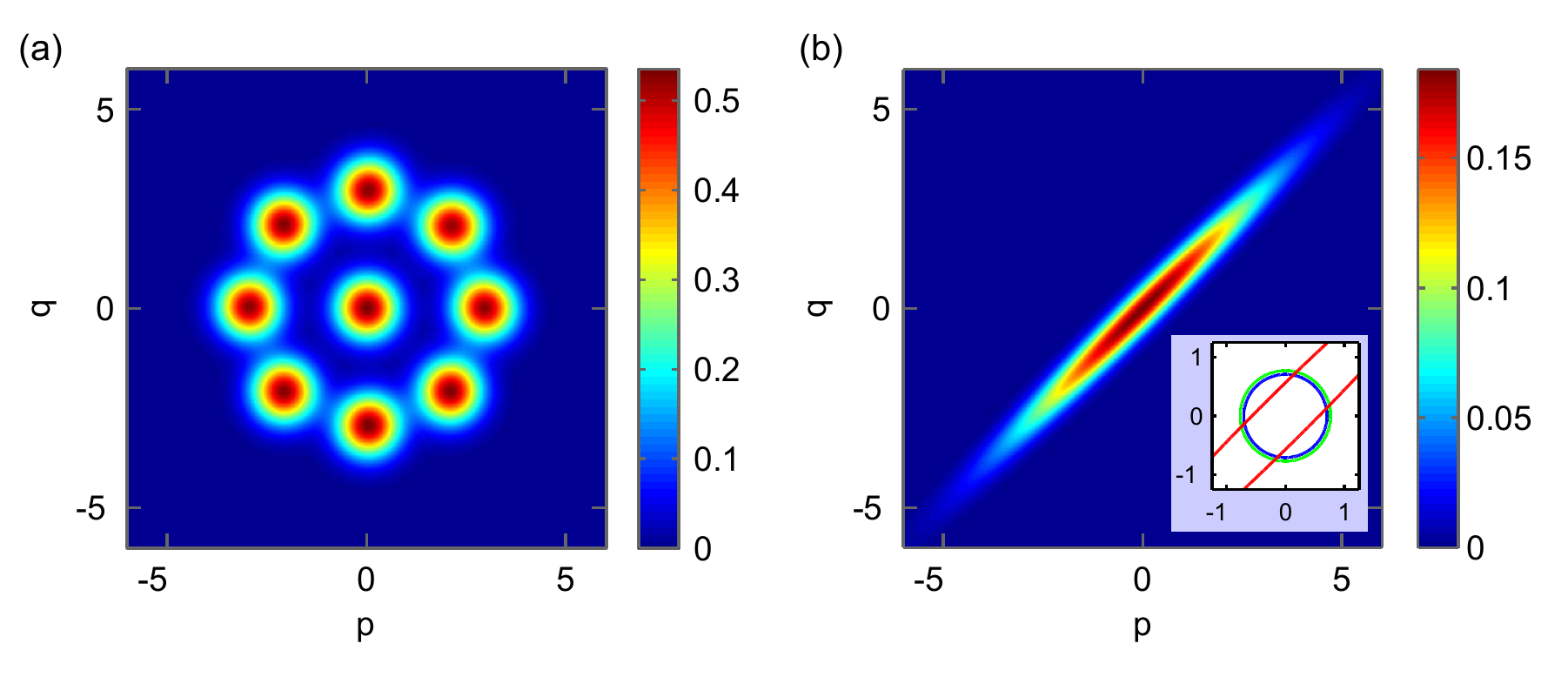}
\caption{\label{Path_Ent_Fig2}{Dual-path reconstruction of various states}
incident at the ``squeezed state input port'' of the beam splitter. $p$ and $q$
are dimensionless variables spanning the phase space. (a) JPA pump off.
Reconstruction of the vacuum and of displaced vacuum states (coherent states,
$8.80{\pm}0.01$~photons, eight different phase values). All nine Wigner
functions are superposed. (b) JPA pump on. Squeezed state for $10$\,dB JPA
signal gain at $45$\degree. Inset:  $1/e$ contours of the ideal vacuum (blue),
the experimental vacuum (green) displayed in panel (a), and the squeezed state
(red).}
\end{figure*}

As a first test of our setup, we perform dual-path reconstructions of the Wigner
function for known input states. Here, we exploit the fact that the noise
contributions of the two amplification paths are independent, while the split
signals are correlated~\cite{Menzel:2010a}(see Supplementary). We reconstruct vacuum
fluctuations and coherent states (displaced vacuum), both at a frequency
$f_0\,{=}\,5.637$\,GHz. Because of narrow-band filtering, we approximate the
vacuum and thermal states as single-mode fields. The results shown in
Fig.~\ref{Path_Ent_Fig2}(a) exhibit a very good phase control for the coherent
state. In addition, we find a small thermal contribution of $0.097{\pm}
0.007$~photons above the vacuum level which can be due to a small thermal
population or other experimental imperfections. In the next step, we generate a
squeezed state by pumping the JPA. For a signal gain of $10$\,dB and a phase of
$45${\degree}, the reconstructed Wigner function is shown in
Fig.~\ref{Path_Ent_Fig2}(b). An analysis of the reconstructed signal moments
reveals that, at the input of the beam splitter, the state generated by the JPA
is squeezed by $4.9{\pm}0.2$\,dB below the vacuum level and contains $8.72{\pm}
0.05$~photons. Furthermore, the product of the standard deviation of the
squeezed quadrature with that of its orthogonal, enlarged one, is $3.45{\pm}
0.07$~times larger than the variance of the ideal vacuum. In other words, we can
model the state as one created by an ideal squeezer acting on an effective
thermal field with $1.22{\pm}0.04$~photons. This thermal field contains the
combined effects of losses and the small thermal population found in the
experimental vacuum.  Again, we notice good control of the phase. It is
noteworthy to mention that the amount of squeezing quoted above is mainly
limited by cable losses and not by the JPA itself.

After characterizing the input fields of the beam splitter, we now turn to its
outputs. With the reference-state method, we build an entanglement witness
matrix from the reconstructed moments. Our witness reliably distinguishes
between ``separable outputs'' for the vacuum state and ``path entangled
outputs'' for the squeezed state input. Next, we analyze the third and fourth
order cumulants and find them to be small for JPA signal gains up to $10$\,dB.
Since this is a strong indication for Gaussian states, we explore the path
entanglement generated in our setup quantitatively via the negativity
$\mathcal{N}_{\rm out}$. For positive values, $\mathcal{N}_{\rm out}$ describes
the degree of entanglement produced between the beam splitter output
paths (see Supplementary). In the limit of low JPA signal gain,
Fig.~\ref{Path_Ent_Fig3}(a) shows how $\mathcal{N}_{\rm out}$ becomes suppressed
when sending more and more thermal photons into the JPA. At some point, the JPA
cannot squeeze the incoming field below the vacuum anymore and the output state
is no longer entangled. For constant temperature, Fig.~\ref{Path_Ent_Fig3}(b)
shows how $\mathcal{N}_{\rm out}$ increases with increasing signal gain from
zero to a value $\mathcal{N}_{\rm out,max}\,{=}\,0.55{\pm}0.04$ at $10$\,dB
signal gain. This behavior is in good agreement with the negativity $\mathcal{N}
_{\rm calc}$ calculated from the dual-path reconstructed input state. Again, we
observe a suppression for large thermal fields sent into the JPA. Our results
confirm the expectation~\cite{Kim:2002a} that the degree of squeezing at the
beam splitter input determines the amount of entanglement generated between the
output paths. However, since $\mathcal{N}_{\rm calc}$ is generally slightly
lower than $\mathcal{N}_{\rm out}$, we conclude that either the dual-path
reconstruction underestimates the squeezing at the beam splitter input or the
reference-state method ignores a small amount of spurious classical correlations
between the two paths. Both effects are consistent with the data shown in
Fig.~\ref{Path_Ent_Fig3}(a), where at constant signal gain, the curve measured
with the reference-state method at the beam splitter output converges for high
temperatures to that calculated from the dual-path reconstructed input state.
We finally note that the path-entangled state is expected to be a two-mode
squeezed state with two additional local squeezing operations applied to
it~\cite{Kim:2002a}. Since local operations do not change the amount of
entanglement, the negativity $\mathcal{N}_{\rm out,max}\,{=}\,0.55{\pm}0.04$
implies that the two-mode squeezed state before the two local operations would
have a variance squeezed by $3.2$\,dB below that of the two-mode vacuum.

\begin{figure*}
\centering
\includegraphics[width=\textwidth]{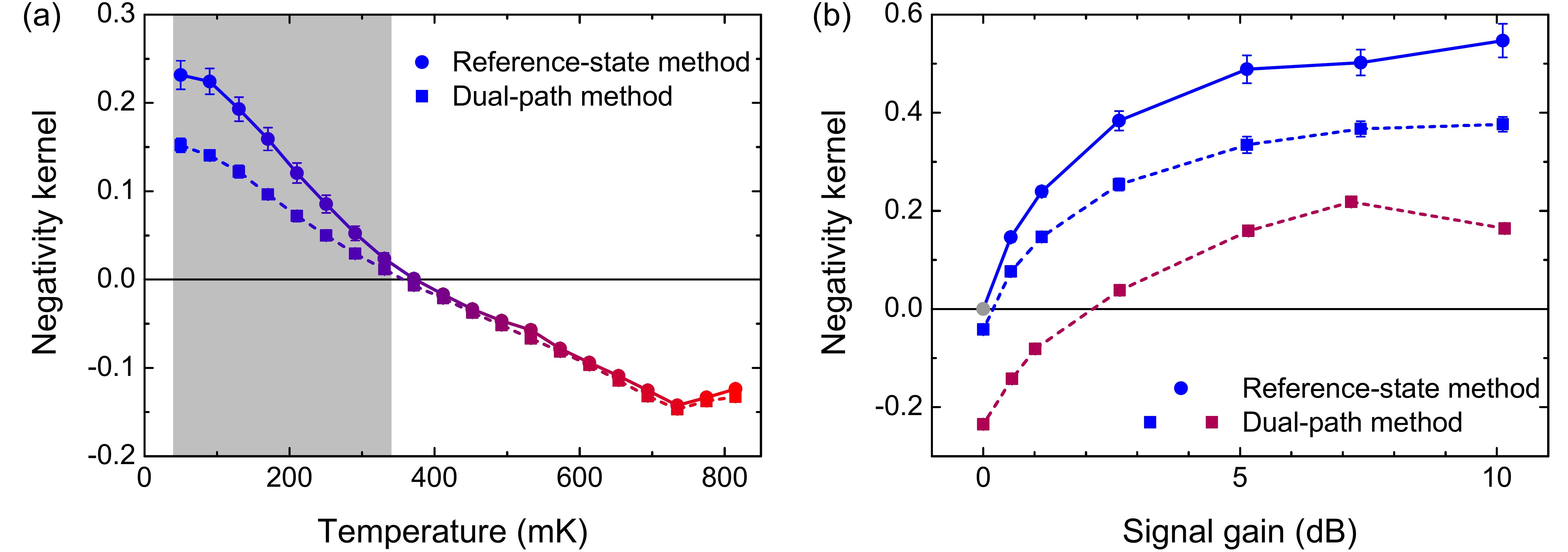}
\caption{\label{Path_Ent_Fig3}{Quantitative analysis of the path entanglement}
generated in our experiments. The negativities $\mathcal{N}_{\rm out}$,
$\mathcal{N}_{\rm calc}$ are the maxima of the corresponding negativity kernels
$\tilde{\mathcal{N}}_{\rm out}$, $\tilde{\mathcal{N}}_{\rm calc}$ and 0.
Circular symbols: $\tilde{\mathcal{N}}_{\rm out}$ data at the beam splitter
output. Square symbols: $\tilde{\mathcal{N}}_{\rm calc}$ calculated from the
reconstructed input state. The lines are guides to the eye. (a) Negativity
kernel versus attenuator temperature (color code) at $1$\,dB signal gain. For
the data points in the shaded area, the witness matrix~\cite{Shchukin:2005a}
confirms entanglement. (b) Negativity kernel versus the JPA signal gain. The
blue (red) curves are recorded at $50$\,mK ($573$\,mK). Grey point: negativity
of the reference state, assumed to be zero.}
\end{figure*}

In summary, we present clear evidence for path entanglement generated by
combining two frequency-degenerate continuous-variable microwave fields, the
vacuum and the squeezed vacuum, in a beam splitter. For an input state squeezed
$4.9{\pm}0.2$\,dB below the vacuum, we observe a maximum negativity $\mathcal{N}
_{\rm out,max}\,{=}\,0.55{\pm}0.04$ at $10$\,dB JPA signal gain. Our experiments
bring the exciting quantum physics of entangled propagating electromagnetic
fields to the technologically highly attractive microwave domain. In this way,
they open up new and exciting perspectives towards microwave quantum
teleportation, quantum communication, and quantum radar~\cite{Lanzagorta:2012}.

The authors thank Christopher Eichler for discussions. We acknowledge support
from the Deutsche Forschungsgemeinschaft via the Sonderforschungsbereich~631,
the German excellence initiative via the `Nanosystems Initiative Munich' (NIM),
from the EU projects SOLID, CCQED and PROMISCE, from MEXT Kakenhi ``Quantum
Cybernetics'', the JSPS through its FIRST Program, the Project for Developing
Innovation Systems of MEXT, the NICT Commissioned Research, EPSRC EP/H050434/1,
Basque Government IT472-10, and Spanish MICINN FIS2009-12773-C02-01.

\end{cbunit}

\clearpage
\begin{cbunit}
\vspace*{0.5cm}
{\noindent \huge \bf Supplementary}
\vspace*{0.5cm}
\setcounter{figure}{0}
\renewcommand{\figurename}{Fig. S\!\!}
\section{Experimental details}

\subsection{The flux driven Josephson parametric amplifier}

We generate squeezed states using a flux driven Josephson parametric amplifier~\cite{Yamamoto:2008a} (JPA). Micrographs of the device used in our experiments are shown in \mbox{Figs.~S\ref{Path_Ent_Sup_Fig1A}(a)--(d)}, a circuit diagram is displayed in Fig.~S\ref{Path_Ent_Sup_Fig1A}(e). 
\begin{figure}[b]
\centering
\includegraphics[width=89mm]{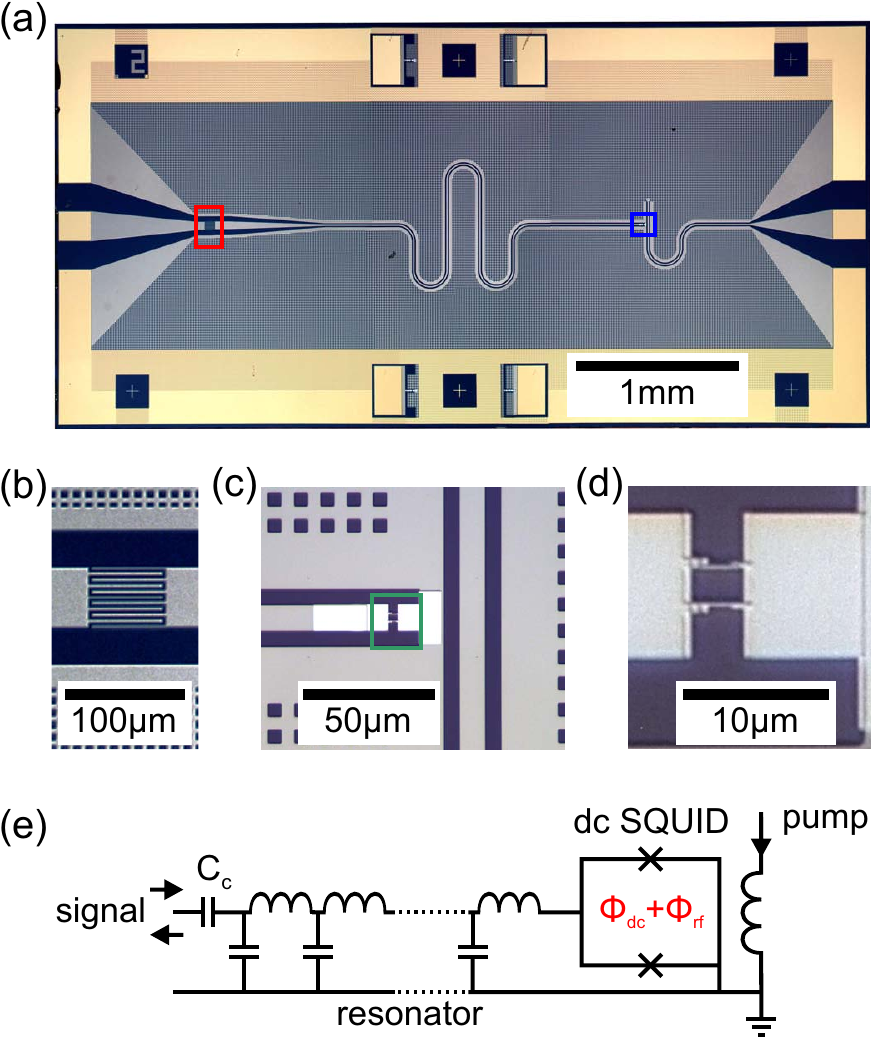}
\caption{\label{Path_Ent_Sup_Fig1A}{Flux driven Josephson parametric amplifier} used in our experiments. (a) Sample chip. (b) Zoom-in of the region marked with the red rectangle in panel (a), showing the coupling capacitor $C_{\rm c}$. (c) Pump line and dc SQUID. Zoom-in of the region marked with the blue rectangle in panel~(a), showing the pump line and the dc SQUID. (d) Zoom-in of the region marked with the green rectangle in panel (c), showing the dc SQUID. (e) Circuit diagram.} 
\end{figure}
In order to achieve a parametric effect, the resonance frequency $f_{\rm dc}$ of a quarter wavelength superconducting coplanar waveguide resonator (external quality $Q_{\rm ext}\,{=}\,312$) is modulated at $2f_{\rm dc}$. To this end, a dc superconducting quantum interference device (SQUID) -- a superconducting loop interrupted by two Josephson junctions -- is inserted between the center conductor and the ground plane at the shorted end of the resonator. 
Because the SQUID acts as a flux-tunable inductor, the resonance frequency of the resonator can be changed by applying an external magnetic field. Via an external coil, we first set a quasistatic bias corresponding to a JPA operating frequency of $f_0\,{=}\,5.637$\,GHz (see Fig.~S\ref{Path_Ent_Sup_Fig1B}). The fast modulation at $2f_0$, also referred to as the pump tone, is then applied in a pulsed fashion (see Sec.~\ref{sec:dataacquisition}) via an on-chip antenna. We determine the isolation between antenna and resonator to be at least $28$\,dB at the signal frequency $f_0$.

\begin{figure}[h]
\centering
\includegraphics[width=89mm]{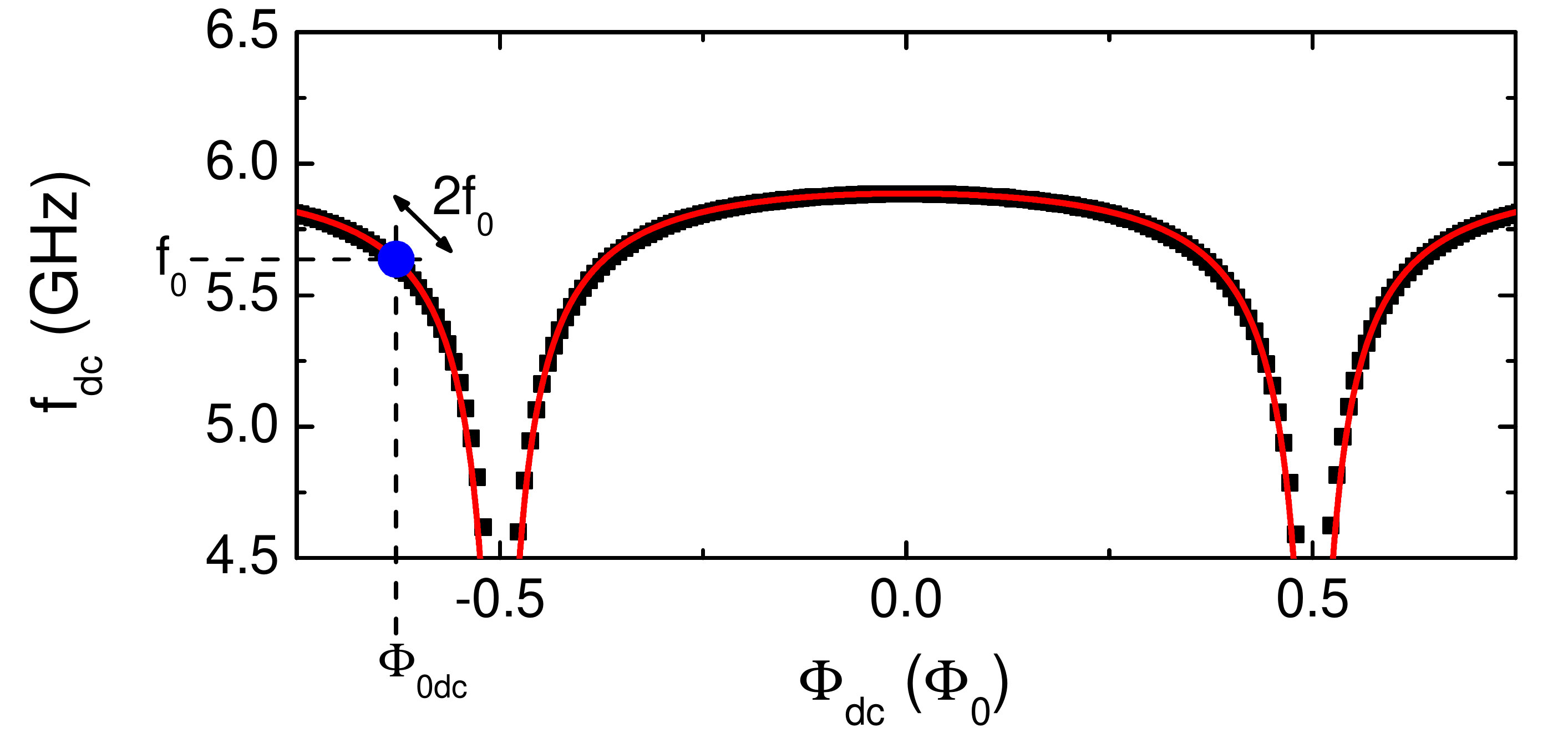}
\caption{\label{Path_Ent_Sup_Fig1B}{JPA resonance frequency} $f_{\rm dc}$ as a function of the applied dc flux $\Phi_{\rm dc}$. Black symbols: data. Red line: fit. Blue dot: operating point $f_0\,{=}\,5.637$\,GHz.}
\end{figure}

Resonator and antenna are made of a $50$\,nm thick Nb film. At the contacts, $95$\,nm of gold on a $5$\,nm titanium bonding layer are deposited on top. As substrate, we use thermally oxidized ($300$\,nm) silicon with a thickness of $300$\,{\micro}m. The dc SQUID is fabricated in the last step using aluminum technology and shadow evaporation. The Al electrodes have a thickness of $50$\,nm each. From Fig.~S\ref{Path_Ent_Sup_Fig1B}, we estimate a Josephson coupling energy $E_{\rm J}/h\,{=}\,650$\,GHz for each junction, where $h\,{=}\,6.63{\times}10^{-34}\,$Js is the Planck constant. The sample chip is placed between two small alumina printed circuit boards inside a gold-plated copper box.

\subsection{The $180 \degree$ hybrid ring microwave beam splitter}

\begin{figure}[h]
\centering
\includegraphics[width=89mm]{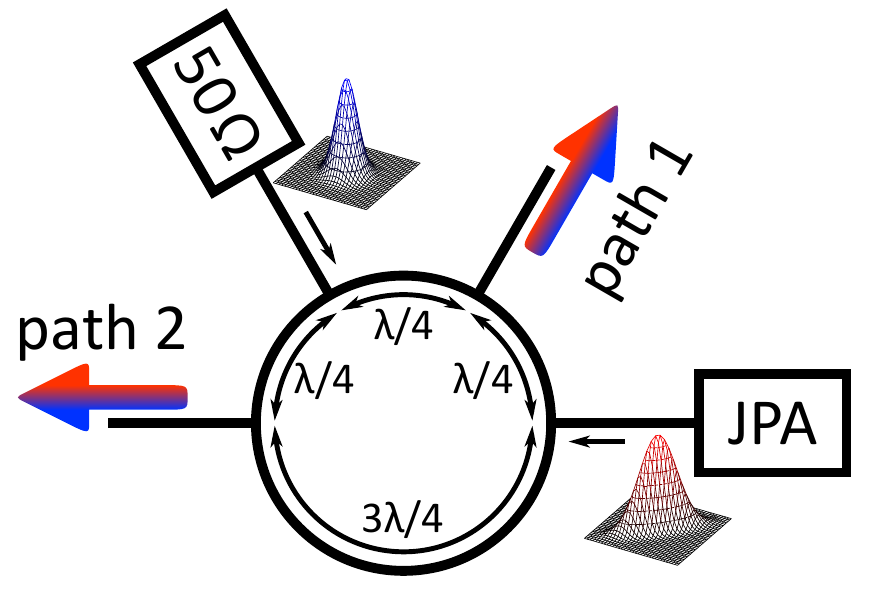}
\caption{\label{Path_Ent_Sup_Fig2}{Schematic sketch of the $180${\degree} hybrid ring microwave beam splitter} used in our experiments. The split signals from the JPA acquire a $180${\degree} phase shift with respect to each other, while the split signals from the 50\,{\ohm} termination remain in phase.} 
\end{figure}

Microwave beam splitters divide an input signal while possibly adding phases to their outputs. However, in order to be lossless, matched, and reciprocal, these devices must necessarily have four ports~\cite{Collin:2001a}. In a quantum mechanical picture, this implies that a second, possibly hidden input port is always present~\cite{Mariantoni:2010a}. A sketch of the 180{\degree} hybrid ring microwave beam splitter used in this work is shown in Fig.~S\ref{Path_Ent_Sup_Fig2}. It is a commercially fabricated device based on gold microstrip transmission lines on a dielectric substrate encased in a copper housing. Its functionality is best understood as follows: at a wavelength $\lambda$, the signals incident at the input ports form an interference pattern in the ring with antinodes at the output ports and nodes at the input ports~\cite{Collin:2001a}. This was experimentally demonstrated in Ref.~\onlinecite{Hoffmann:2010a}, where also typical transmission characteristics are shown. The signals from the two input ports are evenly split and superposed in the two output ports. In this superposition, the component from one of the inputs acquires a $180${\degree} phase shift between the output ports, while that from the other one remains in phase. The center frequency of our hybrid ring is $5.75$\,GHz. Within our measurement bandwidth of $2\,{\times}\,BW_{\rm filter}\,{=}\,2\,{\times}\,489$\,kHz centered at the JPA operating frequency $f_0$ (see also Fig.~S\ref{Path_Ent_Sup_Fig7}), our hybrid ring still has a coupling of $3.5$\,dB between input and output ports and an isolation of at least $38$\,dB between any two input or output ports. The magnitude imbalance between the two output ports is only $0.03$\,dB. For a JPA emitting a squeezed state and a 50\,{\ohm} termination emitting vacuum fluctuations into the input ports of our hybrid ring, the state in the beam splitter output ports is expected to be path-entangled~\cite{Ou:1992a}. 

\subsection{The cross correlation detector}
\label{Path_Ent_Sup_Thecrosscorrelationdetector}

In Fig.~S\ref{Path_Ent_Sup_Fig3}, a simplified sketch of the cross correlation detector is shown. Along each path, the signal emerging from the hybrid ring is linearly amplified, filtered, down-converted to an  intermediate frequency $f_{\rm IF}\,{=}\,11$\,MHz, and digitized at a sampling rate of $150$\,MHz by $16$\,bit analog-to-digital converters. The IQ-mixers used for down-conversion are biased by a strong local oscillator at $5.626$\,GHz and split each output signal into its in-phase ($I_{1,2}$) and quadrature ($Q_{1,2})$ components -- therefore four ADCs are required. The digitized signals are finally fed into an FPGA logic (details can be found in Sec.~\ref{sec:dataacquisition}) which computes all correlations up to the fourth moment in amplitude in real time. Phase synchronization is guaranteed by using a joint local oscillator for down-conversion and referencing the clock of the FPGA logic to that of the local oscillator.

\begin{figure}[h]
\centering
\includegraphics[width=89mm]{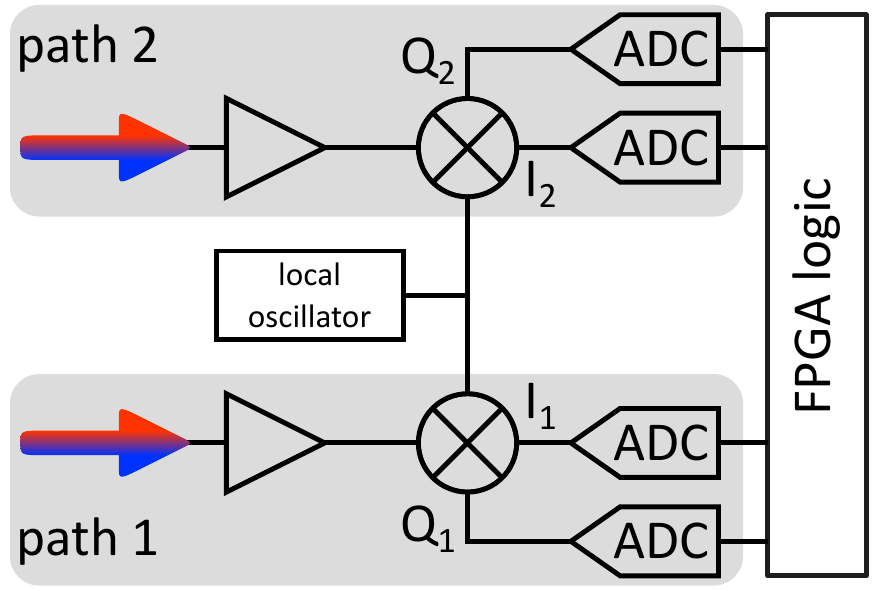}
\caption{\label{Path_Ent_Sup_Fig3}{Sketch of the cross correlation detector.} Colored arrows: output signals from the hybrid ring. Triangular symbols: amplifiers. Circles with crosses: IQ-mixers.} 
\end{figure}

\subsection{Detailed setup}
\label{sec:detailedsetup}

\begin{figure*}[h!]
\centering
\includegraphics[width=0.97\textwidth]{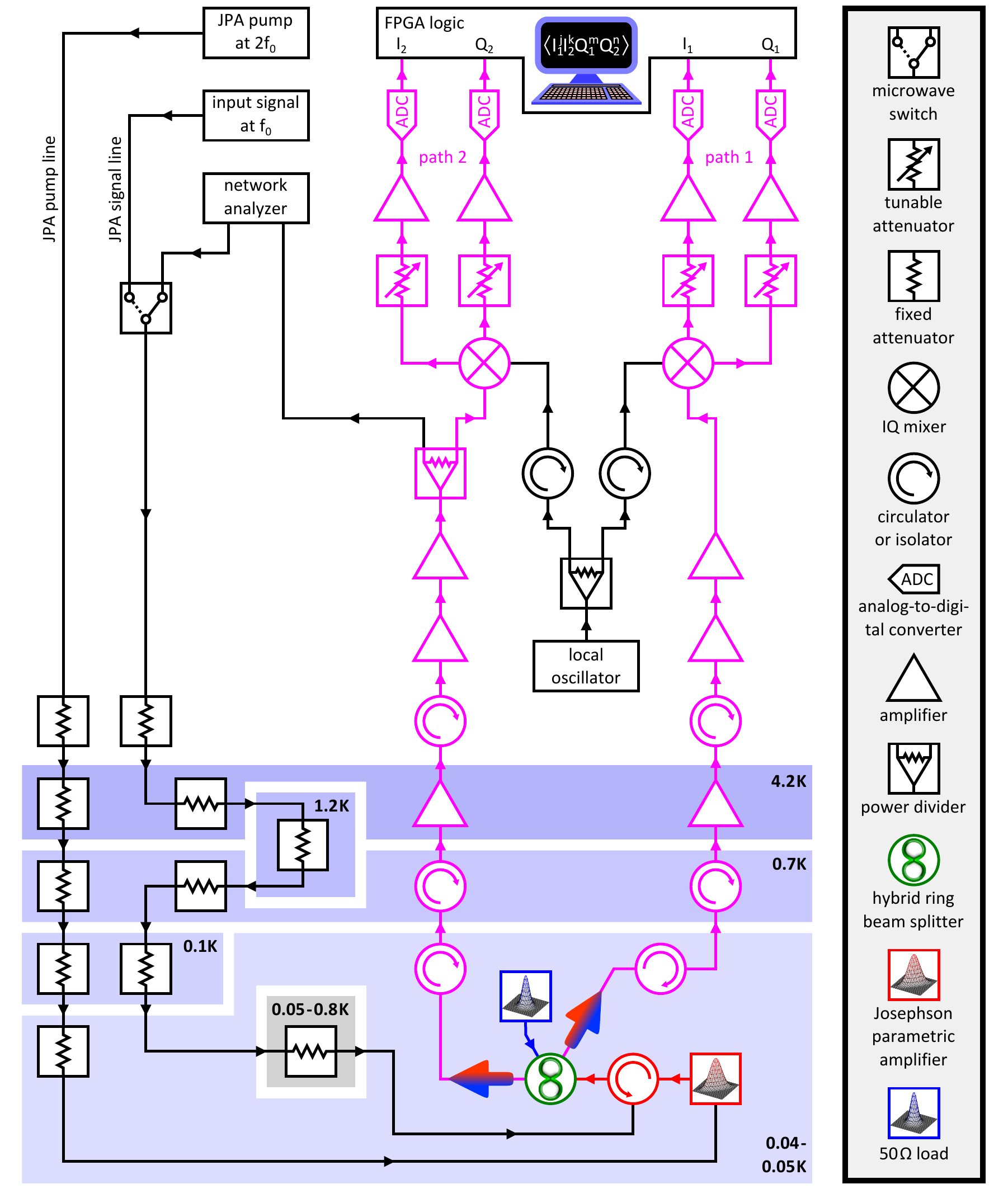}
\caption{\label{Path_Ent_Sup_Fig5}{Detailed experimental setup.} The bold numbers indicate the temperatures of the corresponding colored boxes. The two amplification and detection paths are marked with magenta color. The colored arrows denote the path entangled output state of the beam splitter.}
\end{figure*}

A detailed setup of the experiment is shown in Fig.~S\ref{Path_Ent_Sup_Fig5}. JPA, 50\,{\ohm} terminated hybrid ring, and measurement circulator are anchored to the base temperature plate of a dilution refrigerator, whose temperature is stabilized to $50$\,mK measured on the JPA sample box. Near the 50\,{\ohm} termination of the hybrid ring, we measure a temperature of $40$\,mK. The JPA signal and pump lines are heavily attenuated at various temperature stages. The coldest attenuator of the signal line is only weakly coupled to the base plate and the lower step exchanger of the fridge. Its temperature can therefore be controlled in the range $T_{\rm att}\,{=}\,50\,{-}\,800\,$mK, while all other components retain a stable temperature. This attenuator constitutes a broadband microwave black body emitter, which is used to calibrate the gains of the amplification paths in a Planck spectroscopy experiment~\cite{Mariantoni:2010a}. The total power of each amplification path detected at the ADCs is
\begin{eqnarray}
   P_{1,2}(T_{\rm att}) &=&
   \frac{\langle I_{1,2}^{2} \rangle+\langle Q_{1,2}^{2} \rangle}{R}
   \nonumber\\  
   &=& \frac{\kappa G_{1,2}^{}}{R} \left[\frac{1}{2} \coth\left(\frac{h f_0}{2 k_{\rm B} T_{\rm att}}\right) + n_\mathrm{1,2}^{}\right]\,,
   \label{eqn:calibration}
\end{eqnarray}
where $R\,{=}\,50\,\ohm$ is the input resistance of the ADCs and $k_{\rm B}\,{=}\,1.38{\times}10^{-23}\,\joule/\kelvin$ the Boltzmann constant. The product of the gain $G_{1,2}$ and the photon number conversion factor $\kappa\,{\equiv}\,R{\times}2{\times}BW_{\rm filter}{\times}hf_0\,{=}\,1.83{\times}10^{-16}\,\volt^2$ relates the measured autocorrelations $\langle I_{1,2}^{2}\rangle$ and $\langle Q_{1,2}^{2}\rangle$, which have units of $\volt^2$, to the number of photons of frequency $f_0$ referred to the attenuator. The gain $G_{1,2}$ and the number of noise photons added by each amplification path, $n_{1,2}$, are retrieved from fitting Eq.~(\ref{eqn:calibration}) to the experimentally recorded autocorrelations. For pedagogical reasons, all formulas given in this manuscript assume equal gains and losses for the $I$ and $Q$ branches within each path. However, in the actual evaluation of the data, we do not make this assumption. We model the losses with beam splitters and also account for temperature gradients along our cables. In this way, we make individual fits for the dependence of $\langle I_1^{2}\rangle$, $\langle I_2^{2}\rangle$, $\langle Q_1^{2}\rangle$, and $\langle Q_2^{2}\rangle$ on $T_{\rm att}$. As an example, we show data and fit for $\langle I_1^{2}\rangle$ in Fig.~S\ref{Path_Ent_Sup_Calibration}. We first note that from this figure, we immediately see that the number of thermal photons in the mode $f_0$ is negligible at $40\,{-}\,50$\,mK. Furthermore, with a total loss of $1.8$\,dB between attenuator and beam splitter input, we retrieve $G_{\rm d1}/2\,{=}\,116.5\,\deci\bel$ and $n_\mathrm{d1}\,{=}\,24.3$~photons. Here, the index ``d'' denotes that $G_{\rm d1}/2$ and $n_{\rm d1}$ are referred to the input of the hybrid ring. From our reference-state analysis described in Sec.~\ref{sec:referencestate}, we obtain, with respect to the beam splitter output ports, noise temperatures of $3.00$\,K and $3.27$\,K for the two amplification paths. Considering that our beam splitter reduces the input signal by $3.5$\,dB, the value of $3.00$\,K is in very good agreement with that of $n_{\rm d1}$ quoted above.

\begin{figure}[h]
\centering
\includegraphics[width=89mm]{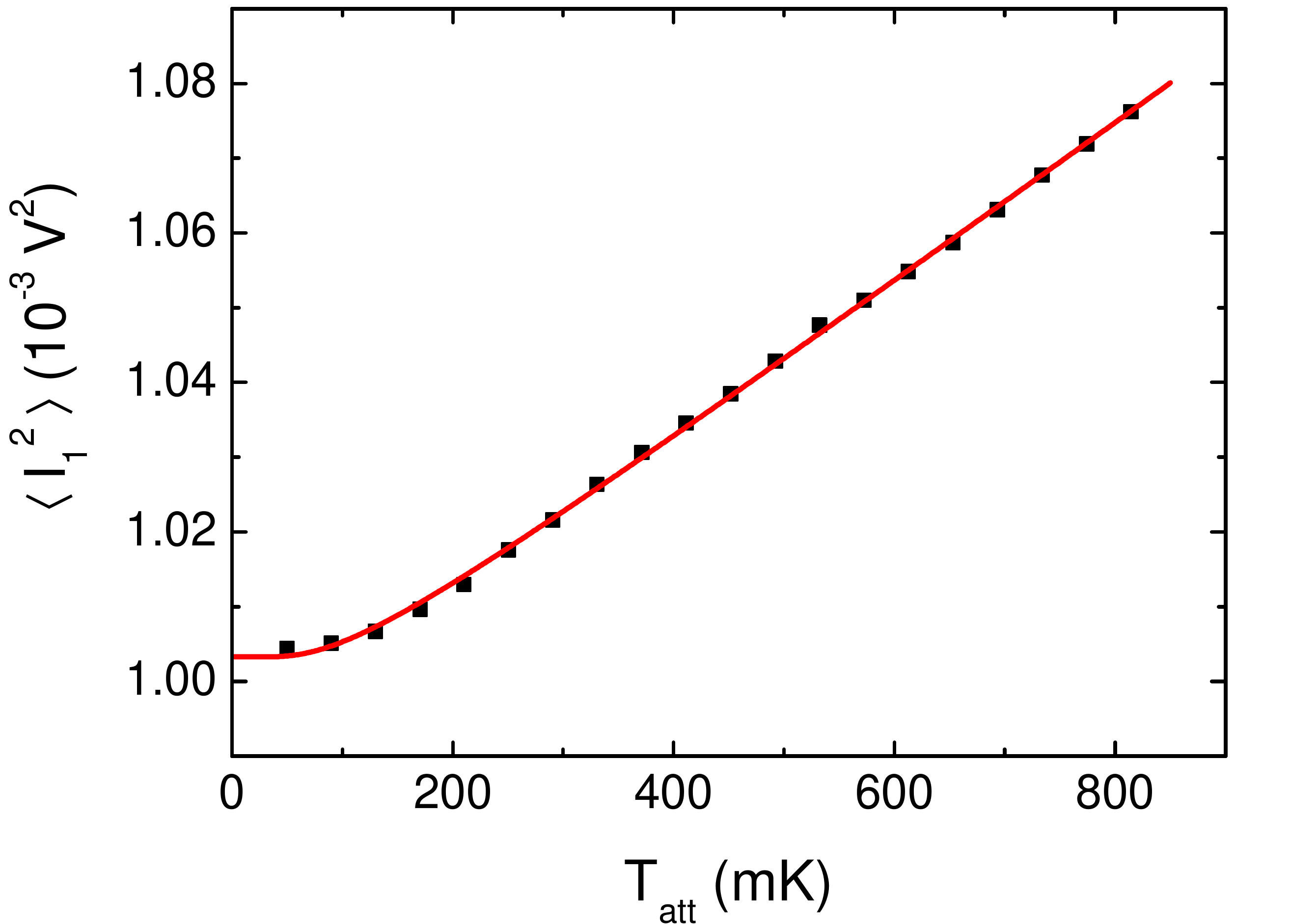}
\caption{\label{Path_Ent_Sup_Calibration}{Gain calibration.} Dependence of the second moment $\langle I_{1}^{2} \rangle$ on the temperature of the attenuator temperature $T_{\rm att}$. Black symbols: data. Red line: fit.} 
\end{figure}

Apart from the key functional elements described in Sec.~\ref{Path_Ent_Sup_Thecrosscorrelationdetector}, the amplification paths contain  isolating elements at various temperature stages to avoid spurious correlated noise contributions. Furthermore, tunable attenuators after the IQ-mixers allow for a prebalancing of the four channels. The JPA pump and signal microwave sources, the local oscillator for the IQ mixers, and the clock for the ADCs and the FPGA logic are synchronized with a Rubidium-based $10$\,MHz source. The vector network analyzer is used to measure the JPA operating point, signal gain, and idler gain. Finally, in addition to the elements shown in Fig.~S\ref{Path_Ent_Sup_Fig5}, several components such as mechanical microwave switches, power dividers, and a spectrum analyzer are used in the real setup. We omit these elements in our discussion here since they serve purely technical purposes (e.g., debugging or switching to other experiments) and are not essential for the functionality of the cross correlation detector and the entanglement detection.

\section{Protocol for cross correlation measurements}

\subsection{Data acquisition}
\label{sec:dataacquisition}

In our experiments, we use either a squeezed state or a coherent state at one of the beam splitter inputs, while a vacuum state is always incident at the second one. During measurements on squeezed states, the JPA pump at frequency $2f_0$ is operated in pulsed mode. Similarly, we also use pulsed probe signals when measuring coherent states of frequency $f_0$. The rise and fall times of the pulse envelopes are approximately $10$\,ns each. As shown in Fig.~S\ref{Path_Ent_Sup_Fig6}, the measurement window always contains an off-region as a reference in addition to the signal. At low enough temperatures, this reference state can be considered as the vacuum (see also Sec.~\ref{sec:referencestate}).

\begin{figure}[h]
\centering
\includegraphics[width=89mm]{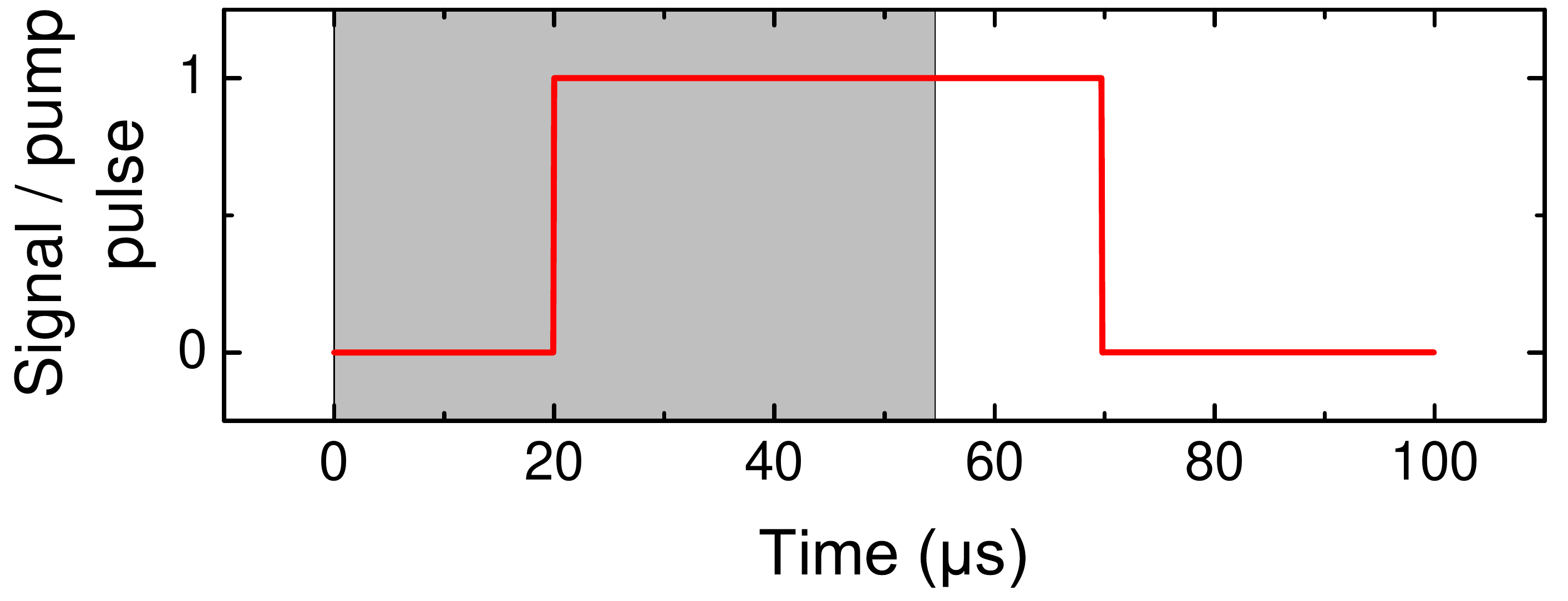}
\caption{\label{Path_Ent_Sup_Fig6}{Measurement protocol} for squeezed and coherent states. During the high time (``1'') of the pulse envelope, either the coherent signal or the JPA pump tone is on, during the low time (``0'') both are off. The shaded area denotes the measurement window.} 
\end{figure}

For the dual-path reconstruction and the entanglement detection, the orthogonal $I$ and $Q$ quadratures of the noisy signal have to be recorded for both paths and products of the type $\langle I_1^jI_2^kQ_1^mQ_2^n\rangle$, where $j\,{+}\,k\,{+}\,m\,{+}\,n\,{\le}\,4$ and $j,k,m,n\,{\in}\,\mathbb{N}_0$, need to be calculated for each recorded data point. However, the presence of the amplifier noise of our paths requires significant averaging. As a consequence, data transfer rate and computation time become a serious bottleneck in a computer-based acquisition system. By streaming the data from the ADCs directly into an FPGA logic, we solve these problems and are able to perform the moment calculations in real time. 

Data acquisition is triggered every $100$\,{\micro}s. For each trigger event, $8192$ consecutive points are digitized in each of the four channels at a rate of $150$\,MHz. As shown in Fig.~S\ref{Path_Ent_Sup_Fig6}, this results in a duty cycle of $54.6$\%. The data is streamed directly into the FPGA, where it first passes a gain balancing and a digital down-conversion stage. The latter also allows for fine-tuning of the phase difference between the two paths to $180${\degree} and for correction for phase imperfections of the IQ-mixer. In the next step, the number of data points is reduced to $512$ with a digital cascaded integrator comb (CIC) filter. The final bandwidth is determined by the subsequent digital finite impulse response (FIR) filter. These filters also determine the measurement bandwidth (noise bandwidth) of $2\,{\times}\,BW_{\rm filter}\,{=}\,2\,{\times}\,489$\,kHz of the cross correlation detector. The corresponding transmission characteristics is shown in Fig.~S\ref{Path_Ent_Sup_Fig7}. 
We note that one of the key advantages of digital filtering is that the frequency dispersion is flat and that the transmission characteristics are exactly equal for all four detection channels. Furthermore, for JPA signal gains up to $20$\,dB, the JPA bandwidth is larger than the measurement bandwidth.

\begin{figure}[h]
\centering
\includegraphics[width=89mm]{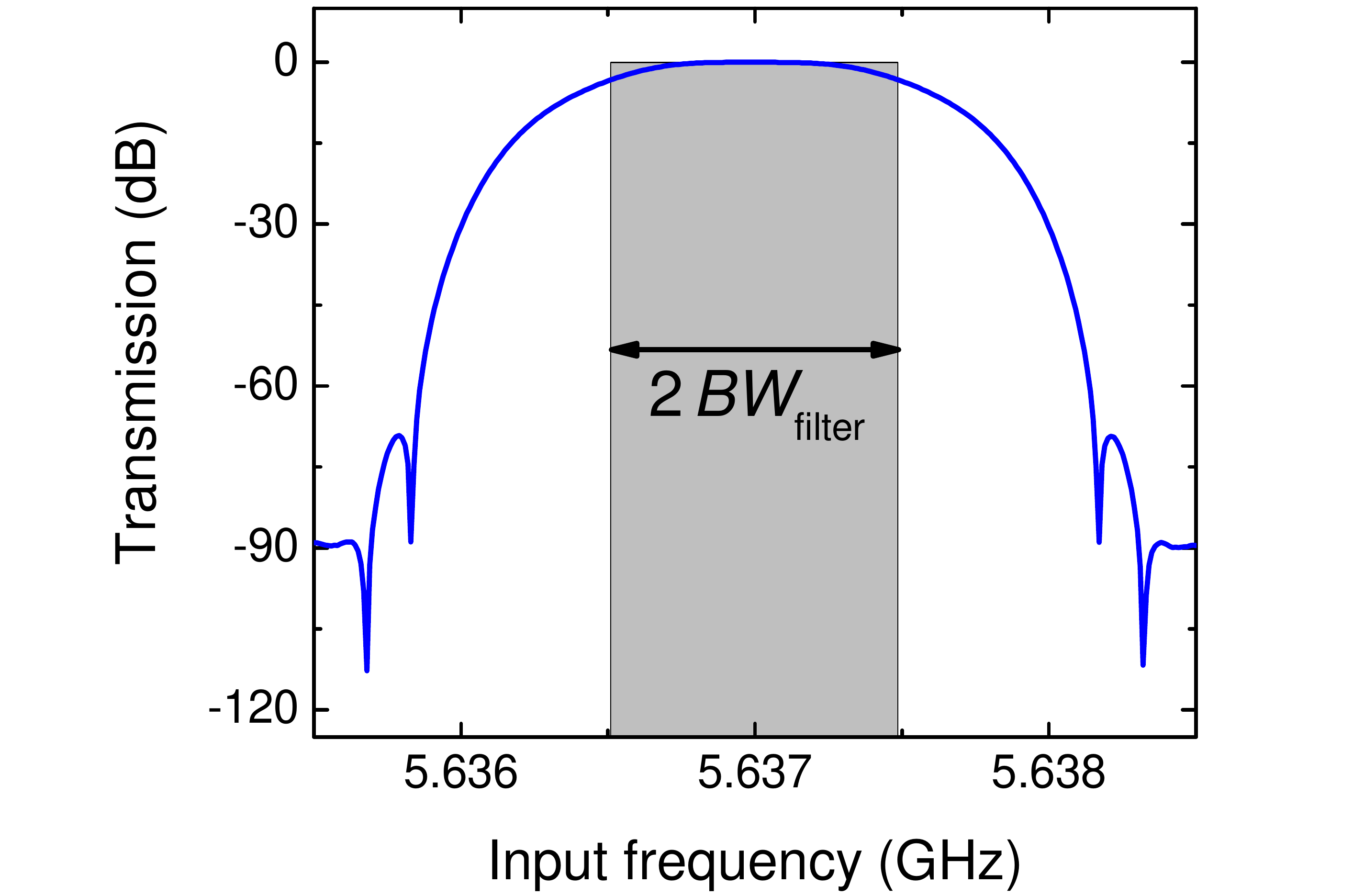}
\caption{\label{Path_Ent_Sup_Fig7}{Measured transmission characteristics} of the cross correlation detector.}
\end{figure}

\begin{figure}[h]
\centering
\includegraphics[width=89mm]{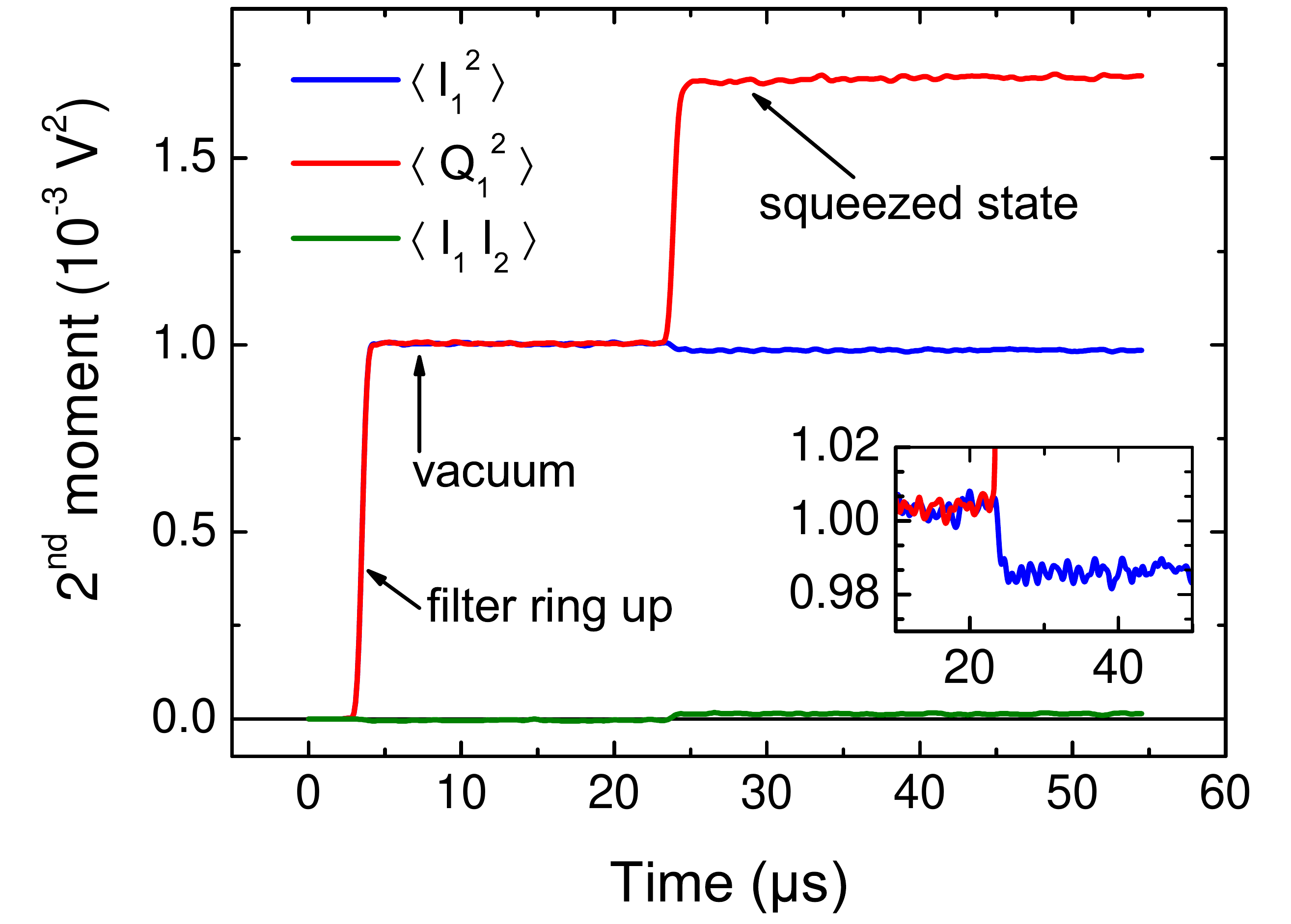}
\caption{\label{Path_Ent_Sup_Fig8}{Typical averaged time traces of selected second moments} from a squeezed state measurement with $0${\degree} phase and for $10$\,dB signal gain. Each averaged trace consists of $5{\times}10^5$ single traces. The rise time of $650$\,ns is determined by the digital filters. The step between JPA pump off (vacuum) and on (squeezed state) is shifted by $4$\,{\micro}s with respect to the pulse shown in Fig.~S\ref{Path_Ent_Sup_Fig6} because of a delay due to filtering.} 
\end{figure}
Finally, the moments up to fourth order in amplitude are calculated. For each moment and each data point, the average over a specified number of trigger events, in the following called ensemble average, is stored inside the FPGA. Figure~S\ref{Path_Ent_Sup_Fig8} shows typical time traces for selected second moments of a squeezed state averaged over $5{\times}10^5$ traces. Since the phase angle in this example is chosen to be $0$\degree, $\langle Q_1^2\rangle$ increases significantly above the vacuum level and $\langle I_1^2\rangle$ decreases below the vacuum level as expected. The cross moment $\langle I_1^{}I_2^{}\rangle$ shows the characteristic cancellation of the uncorrelated noise contributions of the amplification paths, which lies at the heart of the dual-path tomography. A Wigner function reconstruction based on this dataset is shown in Fig.~S\ref{Path_Ent_Sup_Fig9}(b).

\subsection{Dual-path tomography of coherent states}

We first test the dual-path setup against coherent states of frequency $f_0\,{=}\,5.637$\,GHz from a microwave signal generator. The JPA pump is off in these measurements. In Fig.~S\ref{Path_Ent_Sup_Fig9A}, we show the photon number $n\,{\equiv}\,\langle\hat a_{}^\dag\hat a_{}^{}\rangle$ and amplitude $\alpha\,{\equiv}\,\langle\hat a_{}^{}\rangle$ extracted from the reconstructed moments against power $P_{\rm gen}$ at the output of the signal generator. Here, $\hat a_{}^\dag$ and $\hat a_{}^{}$ are the field operators of the input state as defined in Sec.~\ref{sec:dualpath}. The expected linear and square root dependences, $n(P_{\rm gen})\,{=}\,AP_{\rm gen}$ and $\alpha(P_{\rm gen})\,{=}\,B\sqrt{P_{\rm gen}}$, are clearly reproduced. Within an error bar of less than two percent, $B$ is the square root of $A$ for independent fits.

\begin{figure}[b]
\centering
\includegraphics[width=89mm]{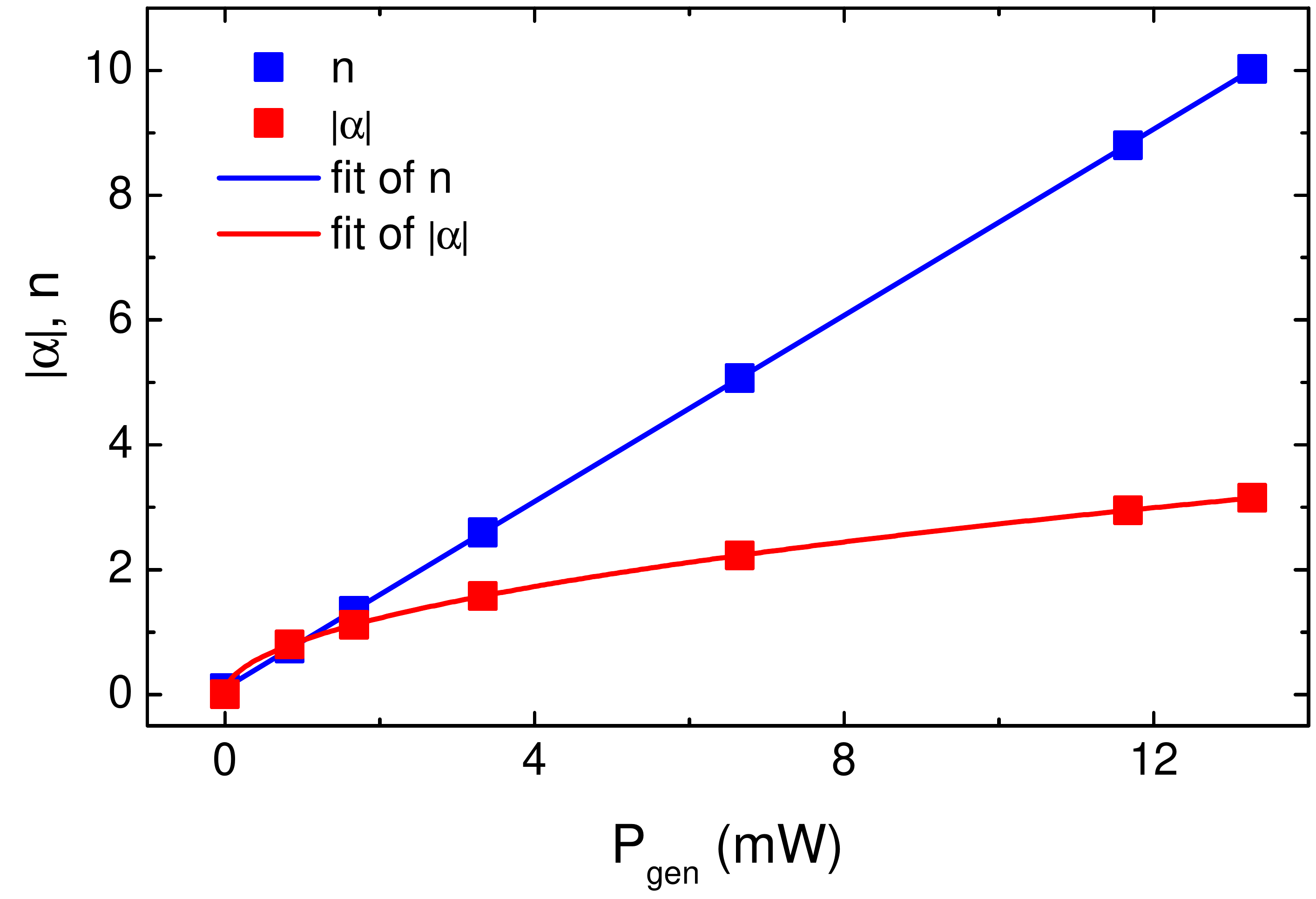}
\caption{\label{Path_Ent_Sup_Fig9A} {Coherent state reconstruction.} Photon number $n$ and amplitude $|\alpha|$ as a function of the signal generator power $P_{\rm gen}$.}
\end{figure}

\subsection{Phase stabilization protocol}

The phase stability between signal/pump and local oscillator is better than $\pm0.3${\degree} for $5{\times}10^5$ traces. Hence, reconstructions based on such a number of averages exhibit very good phase control as shown in Fig.~S\ref{Path_Ent_Sup_Fig9}. However, for quantifying the path entanglement properties, an average over $8{\times}10^6\,{-}\,3{\times}10^7$ traces is necessary to reduce the influence of the noise added by each amplification path to a negligible amount. In such measurements, the phase stability of our setup is not sufficient. For this reason, we record the data in $5{\times}10^5$ trace portions and adjust the relative phase between signal/pump and local oscillator in a way that phase drifts are compensated. In particular, the data shown in Fig.~3 of the main text are recorded in this fashion.

\begin{figure}[h]
\centering
\includegraphics[width=86mm]{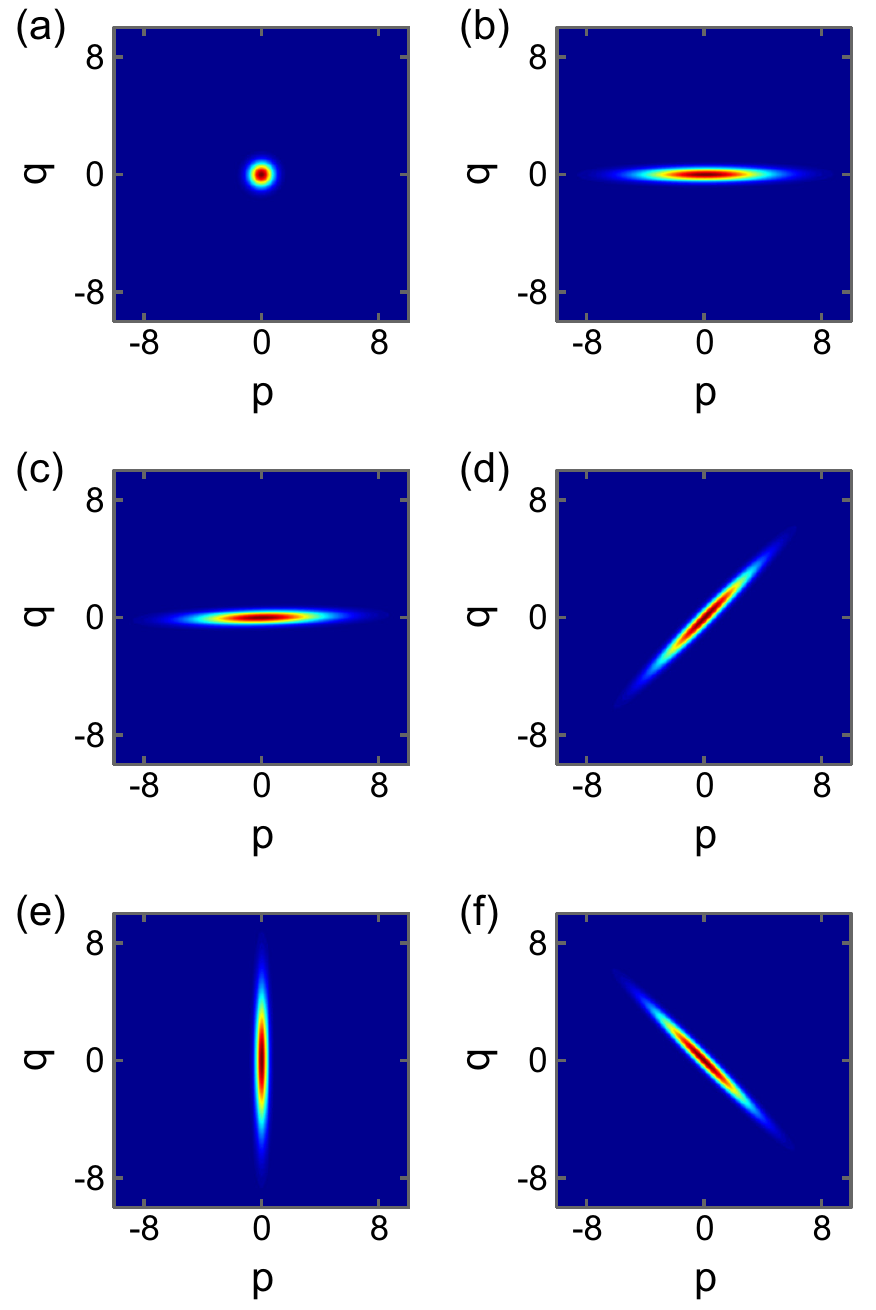}
\caption{\label{Path_Ent_Sup_Fig9}{Phase control for squeezed state reconstruction.} Wigner functions of (a) the vacuum, (b)~a squeezed state at $0$\degree, (c) a squeezed state at $1$\degree, (d) a squeezed state at $45$\degree, (e) a squeezed state at $90$\degree, and (f) a squeezed state at $135$\degree. The number of averaged traces is $5\times10^5$. The residual thermal population of the vacuum is $0.102{\pm}0.005$~photons. For the squeezed state, the JPA signal gain is $10$\,dB and the reconstructed photon number $n\,{=}\,8.67$ varies by approximately $0.5$\% for the different phase angles.}
\end{figure}

\section{Theory: Dual-path tomography and entanglement detection}
In the optical domain, efficient single photon detectors and optical homodyning are established measurement techniques for quantum correlations~\cite{Leonhardt:1997a}. However, despite recent efforts~\cite{Peropadre:2011a,Chen:2012a,Poudel:2012a}, the translation of these methods to the microwave regime remains difficult because of the low photon energy. Therefore, quantum microwave reconstruction based on off-the-shelf noisy linear amplifiers and cross-correlation techniques, the cross-correlation detector~\cite{Menzel:2010a}, was proposed and experimentally implemented~\cite{Menzel:2010a,Bozyigit:2011a}. Later, a second reconstruction technique using only a single amplification path and deconvolution based on a reference measurement was demonstrated~\cite{Eichler:2011a,Eichler:2011b}. In what follows, we describe the mathematical details of these methods, which we a adopted to the needs of our experiments: the reconstruction of the state incident at the input port and the detection of entanglement between signals propagating in the output paths of the beam splitter.

\subsection{Dual-path reconstruction of the input state}
\label{sec:dualpath}

For the input state reconstruction, we can make use of the beam splitter and cross correlations. As described below, we in this way cancel the amplifier noise obscuring the signals because the noise contributions of the two paths are independent. The functionality of microwave beam splitters is well-established for classical signals and was recently demonstrated also for the quantum regime~\cite{Menzel:2010a,Mariantoni:2010a,Bozyigit:2011a}.

For quantum microwaves, we need to take into account the orthogonal signal quadratures $I_{1,2}^{}$ and $Q_{1,2}^{}$. These are real valued voltages measured at the outputs of the IQ-mixers (see Fig.~S\ref{Path_Ent_Sup_Fig3} and Fig.~S\ref{Path_Ent_Sup_Fig5}). We can now define the dimensionless complex envelope functions
\begin{equation}
   \xi_{1,2}^{}\equiv (I_{1,2}^{}+iQ_{1,2}^{})/\sqrt{\kappa}\,,
   \label{eqn:complexenvelope}
\end{equation}
where $\kappa$ is the photon number conversion factor introduced in Sec.~\ref{sec:detailedsetup}.
The corresponding operators,
\begin{equation}
   \hat{\xi}_{1,2}^{}\equiv (\hat{I}_{1,2}^{}+i\hat{Q}_{1,2}^{})/\sqrt{\kappa}\,,
   \label{eqn:complexenvelopeoperator}
\end{equation}
can, in this situation, be expressed as~\cite{DaSilva:2010a}
\begin{equation}\label{S2}
\hat{\xi}_{1,2}^{}=\hat{C}_{1,2}^{}+\hat{v}_{1,2}^\dag\,.
\end{equation} 
Here, $\hat{C}_{1,2}$ is the bosonic annihilation operator of the input signal into the IQ-mixer. The noise added by the latter is represented by its bosonic creation operator $\hat{v}_{1,2}^\dag$. We now have $[\hat{\xi}_{1,2}^{},\hat{\xi}_{1,2}^\dag]\,{=}0$, and the correlations $\langle\hat{\xi}_1^{j^\prime}(\hat{\xi}_1^\dag)_{}^{m^\prime}\hat{\xi}_2^{k^\prime}(\hat{\xi}_2^\dag)_{}^{n^\prime}\rangle\,{=}\,\langle(\hat{\xi}_1^\dag)_{}^{m^\prime}\hat{\xi}_1^{j^\prime}(\hat{\xi}_2^\dag)_{}^{n^\prime}\hat{\xi}_2^{k^\prime}\rangle$ can be expressed in terms of the the measured $\langle I_1^jI_2^kQ_1^mQ_2^n\rangle$ via Eq.~(\ref{eqn:complexenvelope}) by identifying $\hat{\xi}_{1,2}^{}$ with $\xi_{1,2}^{}$ and $\hat{\xi}_{1,2}^\dag$ with $\xi_{1,2}^*$. We note that, while in general, $j^\prime,k^\prime,m^\prime,n^\prime,j,k,m,n\,{\in}\,\mathbb{N}_0$, in this manuscript we restrict ourselves to $j\,{+}\,k\,{+}\,m\,{+}\,n\,{\le}\,4$ or, equivalently, $j^\prime\,{+}\,k^\prime\,{+}\,m^\prime\,{+}\,n^\prime\,{\le}\,4$.

Using the beam splitter relations and the standard quantum model for linear amplifiers~\cite{Caves:1982a}, we can now write
\begin{eqnarray}
   \hat \xi_1^{} &=& \sqrt{\frac{G_{\rm d1}^{}}{2}}({}+\hat a+\hat v)+\sqrt{G_{\rm d1}^{}-1}\hat h_1^\dag+\hat{v}_1^\dag
   \label{eqn:defxione}\\
   \hat \xi_2^{} &=& \sqrt{\frac{G_{\rm d2}^{}}{2}}({}-\hat a+\hat v)+\sqrt{G_{\rm d2}^{}-1}\hat h_2^\dag+\hat{v}_2^\dag
   \label{eqn:defxitwo}
\end{eqnarray}
for our setup. Here, $\hat a_{}^{}$ and $\hat v$ are bosonic annihilation operators. They describe the modes incident on the signal and the 50\,{\ohm} terminated input port of the beam splitter, respectively. The noise fields added by each amplification path are represented by the bosonic creation operators $\hat h_{1,2}^\dag$. Their effective temperature is mainly determined by the noise temperatures of the cold HEMT amplifiers and the cable losses between beam splitter and HEMT amplifier. $G_{\rm d1,2}$ are the gains of the output paths calibrated as described in Sec.~\ref{sec:detailedsetup}. 
In order to simplify the notation, we also define the operators
\begin{eqnarray}
\hat{V}_{1,2}^{}&\equiv&\sqrt{\frac{2}{G_{\rm d1,2}^{}}}\left(\sqrt{G_{\rm d1,2}^{}-1}\,\hat{h}_{1,2}^{}+\hat{v}_{1,2}^{}\right)\label{V}\\
\hat{S}_{1,2}^{}&\equiv&\sqrt{\frac{2}{G_{\rm d1,2}^{}}}\hat{\xi}_{1,2}^{} \label{finalS}\,.
\end{eqnarray}
We note that $\hat V_{1,2}/\sqrt{2}$ is a bosonic operator, as $[\frac{\hat{V}_{1,2}}{\sqrt{2}}, \frac{\hat V_{1,2}^\dag}{\sqrt{2}}]\,{=}\,1$.
In this way, we arrive at the simple expressions
\begin{eqnarray}
   \hat S_1^{} &=&  {}+\hat a+\hat v_{}^{}+\hat V_1^\dag
   \label{eqn:defSone}\\
   \hat S_2^{} &=&  {}-\hat a+\hat v_{}^{}+\hat V_2^\dag
   \label{eqn:defStwo}
  \,.
\end{eqnarray}
We note that the operators $\hat V_{1,2}$, $\hat a$, and $\hat v$, and therefore also $\hat S_{1,2}$, are referred to the input of the beam splitter. With these definitions, we can generalize the dual-path state reconstruction technique, which we originally developed in Ref.~\onlinecite{Menzel:2010a}, in a way that it becomes applicable to the experimental setup presented in this work. Making the reasonable assumptions that $\hat v$ is a weak thermal state with a measured temperature of 40\,mK and that
\begin{equation}
   \langle\hat V_1^{}\rangle =
   \langle\hat V_2^{}\rangle = 0
\end{equation}
for the noise added by the amplification paths, we recursively obtain the signal moments
\begin{widetext}
\begin{align}
&\big\langle(\hat{a}_{}^\dag)_{}^l\hat{a}_{}^{m}\big\rangle_{l_1,m_1}=(-1)_{}^{l-l_1^{}+m-m_1^{}}\big\langle(\hat{S}_1^\dag)_{}^{l_1^{}}(\hat{S}_2^\dag)_{}^{l-l_1^{}}\hat{S}_1^{m_1^{}}\hat{S}_2^{m-m_1^{}}\big\rangle\nonumber\\
&-\sum_{k_1^{}=0}^{l_1^{}}\sum_{k_2^{}=0}^{l-l_1^{}}\sum_{j_1^{}=0}^{m_1^{}}\,\sum_{j_2^{}=0}^{m-m_1^{}-1}\,\,\sum_{k^\prime_1=0}^{l_1^{}-k_1^{}}\,\,\sum_{k^\prime_2=0}^{l-l_1^{}-k_2^{}}\,\,\sum_{j^\prime_1=0}^{m_1^{}-j_1^{}}\,\,\sum_{j^\prime_2=0}^{m-m_1^{}-j_2^{}}\binom{l_1^{}}{k_1^{}}\binom{l-l_1^{}}{k_2^{}}\binom{m_1^{}}{j_1^{}}\binom{m-m_1^{}}{j_2^{}}\nonumber\\
&\quad\times\binom{l_1^{}-k_1^{}}{k^\prime_1}\binom{l-l_1^{}-k_2^{}}{k^\prime_2}\binom{m_1^{}-j_1^{}}{j^\prime_1}\binom{m-m_1^{}-j_2^{}}{j^\prime_2}(-1)_{}^{l-l_1^{}+m-m_1^{}+j_2^{}+k_2^{}}\nonumber\\
&\quad\times\big\langle(\hat{a}_{}^\dag)^{k_1^{}+k_2^{}}\,\hat{a}_{}^{j_1^{}+j_2^{}}\big\rangle\big\langle(\hat{v}_{}^\dag)_{}^{k^\prime_1+k^\prime_2}\,\hat{v}_{}^{j^\prime_2+j^\prime_1}\big\rangle\big\langle\hat{V}_1^{l_1^{}-k_1^{}-k^\prime_1}\,(\hat{V}_1^\dag)^{m_1^{}-j_1^{}-j^\prime_1}\big\rangle\big\langle\hat{V}_2^{l-l_1^{}-k_2^{}-k^\prime_2}(\hat{V}_2^\dag)^{m-m_1^{}-j_2^{}-j^\prime_2}\big\rangle\nonumber\\
&-\sum_{k_1^{}=0}^{l_1^{}}\sum_{k_2^{}=0}^{l-l_1^{}}\,\sum_{j_1^{}=0}^{m_1^{}-1}\,\,\sum_{k^\prime_1=0}^{l_1^{}-k_1^{}}\,\,\sum_{k^\prime_2=0}^{l-l_1^{}-k_2^{}}\,\,\sum_{j^\prime_1=0}^{m_1^{}-j_1^{}}\binom{l_1^{}}{k_1^{}}\binom{l-l_1^{}}{k_2^{}}\binom{m_1^{}}{j_1^{}}\binom{l_1^{}-k_1^{}}{k^\prime_1}\binom{l-l_1^{}-k_2^{}}{k^\prime_2}\binom{m_1^{}-j_1^{}}{j^\prime_1}\nonumber\\
&\quad\times(-1)_{}^{l-l_1^{}+k_2^{}}\big\langle(\hat{a}_{}^\dag)_{}^{k_1^{}+k_2^{}}\,\hat{a}_{}^{j_1^{}+m-m_1^{}}\big\rangle\big\langle(\hat{v}_{}^\dag)_{}^{k^\prime_1+k^\prime_2}\,\hat{v}_{}^{j^\prime_1}\big\rangle\big\langle\hat{V}_1^{l_1^{}-k_1^{}-k^\prime_1}\,(\hat{V}_1^\dag)^{m_1^{}-j_1^{}-j^\prime_1}\big\rangle\big\langle\hat{V}_2^{l-l_1^{}-k_2^{}-k^\prime_2}\big\rangle\nonumber\\
&-\sum_{k_1^{}=0}^{l_1^{}}\sum_{k_2^{}=0}^{l-l_1^{}-1}\binom{l_1^{}}{k_1^{}}\binom{l-l_1^{}}{k_2^{}}(-1)_{}^{l-l_1^{}+k_2^{}}\big\langle(\hat{a}_{}^\dag)_{}^{k_1^{}+k_2^{}}\,\hat{a}_{}^{m}\big\rangle\big\langle\hat{V}_1^{l_1^{}-k_1^{}}\big\rangle\big\langle\hat{V}_2^{l-l_1^{}-k_2^{}}\big\rangle\nonumber\\
&-\sum_{k_1^{}=0}^{l_1^{}-1}\binom{l_1^{}}{k_1^{}}\big\langle(\hat{a}_{}^\dag)_{}^{k_1^{}+l-l_1^{}}\,\hat{a}_{}^{m}\big\rangle\big\langle\hat{V}_1^{l_1^{}-k_1^{}}\big\rangle
\label{eqn:dualpathmoments}
\end{align}

for $l,m,l_1^{},m_1^{}\,{\in}\,\mathbb{N}_0$ from the measured noisy correlations. In this process, we also have to compute the noise moments
\begin{align}
&\big\langle\hat{V}_1^r(\hat{V}_1^\dag)_{}^s\big\rangle=\big\langle(\hat{S}_1^\dag)_{}^r\hat{S}_1^s\big\rangle\nonumber\\
&-\sum_{k_1^{}=0}^{r}\,\sum_{j_1^{}=0}^{s-1}\,\sum_{k^\prime_1=0}^{r-k_1}\,\sum_{j^\prime_1=0}^{s-j_1^{}}\binom{r}{k_1^{}}\binom{s}{j_1^{}}\binom{r-k_1^{}}{k^\prime_1}\binom{s-j_1^{}}{j^\prime_1}\big\langle(\hat{a}_{}^\dag)_{}^{k^\prime_1}\,\hat{a}_{}^{j^\prime_1}\big\rangle\big\langle(\hat{v}_{}^\dag)^{r-k_1^{}-k^\prime_1}\,\hat{v}_{}^{s-j_1^{}-j^\prime_1}\big\rangle\big\langle\hat{V}_1^{k_1^{}}(\hat{V}^\dag_1)^{j_1^{}}\big\rangle\nonumber\\
&-\sum_{k_1^{}=0}^{r-1}\binom{r}{k_1^{}}\big\langle(\hat{a}_{}^\dag)^{r-k_1^{}}\big\rangle\big\langle\hat{V}_1^{k_1^{}}(\hat{V}^\dag_1)_{}^{s}\big\rangle,\label{V1}\\
\nonumber\\
&\big\langle\hat{V}_2^r(\hat{V}_2^\dag)_{}^s\big\rangle=\big\langle(\hat{S}_2^\dag)_{}^r\hat{S}_2^s\big\rangle\nonumber\\
&-\sum_{k_1^{}=0}^{r}\,\sum_{j_1^{}=0}^{s-1}\,\sum_{k^\prime_1=0}^{r-k_1^{}}\,\sum_{j^\prime_1=0}^{s-j_1^{}}\binom{r}{k_1^{}}\binom{s}{j_1^{}}\binom{r-k_1^{}}{k^\prime_1}\binom{s-j_1^{}}{j^\prime_1}(-1)_{}^{k^\prime_1+j^\prime_1}\big\langle(\hat{a}_{}^\dag)^{k^\prime_1}\,\hat{a}_{}^{j^\prime_1}\big\rangle\big\langle(\hat{v}_{}^\dag)^{r-k_1^{}-k^\prime_1}\,\hat{v}_{}^{s-j_1^{}-j^\prime_1}\big\rangle\big\langle\hat{V}_2^{k_1^{}}(\hat{V}^\dag_2)^{j_1^{}}\big\rangle\nonumber\\
&-\sum_{k_1^{}=0}^{r-1}\binom{r}{k_1^{}}(-1)_{}^{r-k_1^{}}\big\langle(\hat{a}_{}^\dag)^{r-k_1^{}}\big\rangle\big\langle\hat{V}_2^{k_1^{}}(\hat{V}^\dag_2)^{s}\big\rangle\label{V2}
\end{align}
associated with both amplification paths for $r\,{+}\,s\,{>}\,1$ and $r,s\,{\in}\,\mathbb{N}_0$, again in a recursive fashion.
\end{widetext}
In other words, the formulas for the moments of order $l\,{+}\,m$ are established using those of the moments of order $l\,{+}\,m\,{-}\,1$. The formulas obtained in this way are not unique, they depend on the specific choices of $l_1^{}$ and $m_1^{}$. We find that the statistical uncertainty in our results is minimized by using the mean value of all formulas found for constant $l\,{+}\,m$.

From the reconstructed signal moments of Eq.~(\ref{eqn:dualpathmoments}), we can readily extract the squeezing below the vacuum in decibel,
\begin{multline}
10\log_{10}\big(-\langle\hat a_{}^2\rangle{\rm e}_{}^{-i\phi}-\langle(\hat a_{}^\dag)_{}^2\rangle{\rm e}_{}^{i\phi}+2\langle\hat a_{}^\dag\hat a\rangle+1 \\\shoveleft\qquad\qquad
 +\langle\hat a\rangle_{}^2{\rm e}_{}^{-i\phi}+\langle\hat a_{}^\dag\rangle_{}^2{\rm e}_{}^{i\phi}-2\langle\hat a_{}^\dag\rangle\langle\hat a\rangle\big)
\,.
\end{multline}
Here, the angle $\phi$ is defined via the relation $\langle\hat a_{}^2\rangle\,{-}\,\langle\hat a\rangle^2\,{=}\,|\langle\hat a_{}^2\rangle\,{-}\,\langle\hat a\rangle^2|{\rm e}_{}^{i\phi}$ and the argument of the logarithm is the ratio between the variance of the squeezed quadrature and the vacuum variance.

In the case of infinitely many reconstructed moments $\langle(\hat a_{}^\dag)_{}^l\hat a_{}^m\rangle$, the Wigner function $W(q,p)$ of an arbitrary state can be completely reconstructed. However, in this work we record these moments only up to fourth order, $l\,{+}\,m\,{\le}\,4$. As we find that the higher moments are consistent with those of Gaussian states, we are allowed to restrict ourselves to moments with $l\,{+}\,m\,{\le}\,2$. This enables us to use an analytical approach~\cite{Buzek:1996a,Buzek:1996b}, which yields
\begin{widetext}
\begin{equation}      
 W(q,p)=\frac{1}{\pi \sqrt{(\nu+1/2)^2-|\mu|^2}}\exp\left[-\frac{(\nu+1/2)|\zeta -\langle \hat a \rangle|^2-(\mu^{*}/2)(\zeta-\langle \hat a \rangle)^2-(\mu/2)(\zeta^{*}-\langle \hat a^\dag \rangle)^2}{(\nu+1/2)^2-|\mu|^2} \right],     
   \label{eqn:Wignerref}
\end{equation}
\end{widetext}
with $\zeta\,{\equiv}\,q\,{+}\,ip$, $\mu\,{\equiv}\,\langle\hat{a}^2\rangle\,{-}\,\langle\hat{a}\rangle^2$, and $\nu\,{\equiv}\,\langle\hat a^\dag\hat a\rangle\,{-}\,|\langle \hat a \rangle|^2$. As explained in Sec.~\ref{sec:dualpath}, we have chosen our definitions such that phase space variables $q$ and $p$ are dimensionless and their value represents the square root of a photon number. Since any Gaussian state can be written as a displaced squeezed thermal state, we can also extract the effective mode temperature analytically from the reconstructed input state moments~\cite{Buzek:1996a,Buzek:1996b}. We note that this effective temperature contains contributions from the physical temperature and from losses.

\subsection{Reference-state analysis of the output state moments}
\label{sec:referencestate}

In order to detect the entanglement between the two paths independently from the dual-path reconstruction of the input state, we cannot assume that the hybrid ring is working as a beam splitter. We therefore follow a different route: we reconstruct the moments of the output state by means of a calibration against a well-known reference signal~\cite{Eichler:2011a,Eichler:2011b}. The obvious choice for this reference signal is the two-mode vacuum. In this way, the beam splitter is treated as a black box device which, for a vacuum state at each input, produces uncorrelated vacuum states at each output port. This rather general assumption holds well for the temperatures measured for attenuator and termination, $40\,{-}\,50$\,mK. In this situation, the complex envelope operator becomes
\begin{eqnarray}
   \hat \xi_1^{} &=&  \sqrt{G_{\rm r1}^{}}\hat s_1^{}+\sqrt{G_{\rm r1}^{}-1}\hat h_1^\dag+\hat{v}_1^\dag
   \label{eqn:defxioneref}\\
   \hat \xi_2^{} &=&  \sqrt{G_{\rm r2}^{}}\hat s_2^{}+\sqrt{G_{\rm r2}^{}-1}\hat h_2^\dag+\hat{v}_2^\dag\,.
   \label{eqn:defxitwo1ref}
\end{eqnarray}
Here, $\hat s_{1,2}^{}$ is referred to the output of the beam splitter, and $G_{\rm r1,2}$ is the effective gain of the amplification paths. Note that the $G_{\rm r1,2}$ are numerically different from $G_{\rm d1,2}$ because they do not contain the beam splitter losses. After defining the operators
\begin{eqnarray}
\hat{V}_{1,2}^{}&\equiv& \sqrt{\frac{1}{G_{\rm r1,2}^{}}}\left(\sqrt{G_{\rm r1,2}^{}-1}\,\hat{h}_{1,2}^{}+\hat{v}_{1,2}^{}\right)\label{Vref}\\
\hat{S}_{1,2}^{}&\equiv& \sqrt{\frac{1}{G_{\rm r1,2}^{}}}\hat{\xi}_{1,2}^{} \label{finalSref}\,,
\end{eqnarray}
we again arrive at the simplified expressions
\begin{eqnarray}
   \hat S_1^{} &=&  \hat s_1^{}+\hat V_1^\dag
   \label{eqn:defSoneref}\\
   \hat S_2^{} &=&  \hat s_2^{}+\hat V_2^\dag
   \label{eqn:defStworef}
  \,.
\end{eqnarray}
We now evaluate the correlations of the outputs of the two channels,
\begin{align}
   &\label{eqn:scorr}
   \big\langle(\hat S_1^\dag)_{}^{l_1^{}}\hat S_1^{m_1^{}}
   (\hat S_2^\dag)_{}^{l_2}\hat S_2^{m_2}\big\rangle 
   \nonumber\\
   &\quad=\big\langle(\hat s_1^\dag+\hat V_1^{})_{}^{l_1^{}}
    (\hat s_1^{}+\hat V_1^\dag)_{}^{m_1^{}}
    (\hat s_2^\dag+\hat V_2{})_{}^{l_2^{}}
    (\hat s_2^{}+\hat V_2^\dag)_{}^{m_2^{}}
    \big\rangle
    \nonumber\\
   &\quad=
    \sum_{k_1^{}=0}^{l_1^{}}\,\sum_{k_2^{}=0}^{l_2^{}}
    \,\sum_{j_1^{}=0}^{m_1^{}}\,\sum_{j_2^{}=0}^{m_2^{}}
    \binom{l_1^{}}{k_1^{}}
    \binom{l_2^{}}{k_2^{}}
    \binom{m_1^{}}{j_1^{}}
    \binom{m_2^{}}{j_2^{}}
    \nonumber\\
    &\quad\quad\times
    \big\langle
    (\hat s_1^\dag)_{}^{l_1^{}-k_1^{}}\hat s_1^{m_1^{}-j_1^{}}
    (\hat s_2^\dag)_{}^{l_2^{}-k_2^{}}\hat s_2^{m_2^{}-j_2^{}}
    \big\rangle
    \nonumber\\
   &\quad\quad\times
    \big\langle
    \hat V_1^{k_1^{}}(\hat V_1^\dag)_{}^{j_1^{}}
    \hat V_2^{k_2^{}}(\hat V_2^\dag)_{}^{j_2^{}}
    \big\rangle\,.
\end{align}
With the terms $\langle(\hat s_1^\dag)_{}^{l_1^{}-k_1^{}}\hat s_1^{m_1^{}-j_1^{}}(\hat s_2^\dag)_{}^{l_2^{}-k_2^{}}\hat s_2^{m_2^{}-j_2^{}}\rangle$, which can be calculated straightforwardly for our reference state, Eq.~(\ref{eqn:scorr}) forms a system of linear equations. The latter allows us to extract the noise terms $\langle\hat V_1^{k_1^{}}(\hat V_1^\dag)_{}^{j_1^{}}\hat V_2^{k_2^{}}(\hat V_2^\dag)_{}^{j_2^{}}\rangle$ related to our amplification paths by algebraic inversion.  Once knowing these noise terms, we can extract the signal correlations for the squeezed state input again from Eq.~(\ref{eqn:scorr}) and algebraic inversion. We note that also more sophisticated reference states and device models, such as thermal states incident at a beam splitter, can be chosen as reference state. Depending on the pre-characterization of the used components, this approach might account better for experimental imperfections. Although the latter typically tend to reduce the degree of entanglement, our entanglement detection  turns out to be quite robust against them.

Altogether, the method described above would, in principle, allow for a reconstruction of the output state as far as this is possible with four moments. We note that from such a reconstruction also all entanglement properties could be derived. However, as shown in the next section we choose a different approach.

\subsection{Entanglement witness}

The detection of entanglement requires substantially less information than a full state reconstruction. Indeed, there exists an infinite number of witnesses and criteria which allow one to decide whether or not a state is entangled. Each of these criteria uses only a small amount of information about the examined state. In this work, we use the witness matrix~\cite{Shchukin:2005a}
\begin{equation}
{\bf M_{}^{(2)}}\equiv
\begin{pmatrix}
1 & \langle   \hat s_1^{}  \rangle & \langle   \hat s_1^\dag \rangle & \langle   \hat s_2^\dag \rangle & \langle   \hat s_2^{} \rangle 
\\ 
\langle\hat s_1^\dag \rangle & \langle   \hat s_1^\dag \hat s_1^{} \rangle & \langle   (\hat s_1^\dag)^{2} \rangle & \langle   \hat s_1^\dag \hat s_2^\dag \rangle & \langle   \hat s_1^\dag \hat s_2^{} \rangle 
\\ 
\langle\hat s_1^{} \rangle & \langle   \hat s_1^{2} \rangle &    1 + \langle \hat s_1^\dag \hat s_1^{} \rangle & \langle   \hat s_1^{} \hat s_2^\dag \rangle & \langle   \hat s_1^{} \hat s_2^{} \rangle 
\\ 
\langle\hat s_2^{} \rangle & \langle   \hat s_1^{} \hat s_2^{} \rangle & \langle   \hat s_1^\dag \hat s_2^{} \rangle & \langle   \hat s_2^\dag \hat s_2^{} \rangle & \langle   \hat s_2^{2} \rangle 
\\ 
\langle\hat s_2^\dag \rangle & \langle   \hat s_1^{} \hat s_2^\dag \rangle & \langle   \hat s_1^\dag \hat s_2^\dag \rangle & \langle   (\hat s_2^\dag)^{2} \rangle &    1 + \langle \hat s_2^\dag \hat s_2^{} \rangle
\end{pmatrix}\,,
\end{equation}
which contains up to second order moments of the beam splitter output state. If $\bf M_{}^{(2)}$ has at least one negative eigenvalue, the state is entangled. The absence of a negative eigenvalue implies separability only in the case of Gaussian states.

\subsection{Negativity}
\label{sec:negativity}

For a bipartite system, the amount of entanglement between the subsystems A and B can be quantified by means of the negativity
\begin{equation}
   {\cal N}(\rho)\equiv\frac{||\rho_{}^{T_B}||_1^{}-1}{2}\,,
\end{equation}
where $\rho$ is the density matrix of the total system, and $||\rho_{}^{T_B}||_1^{}=Tr|\rho_{}^{T_B}|$ is the trace norm of the partial transpose of $\rho$ with respect to subsystem B, $\rho_{}^{T_B}$. If ${\cal N}(\rho)\,{>}\,0$, the state is entangled. For a maximally entangled state, ${\cal N}(\rho)\,{\rightarrow}\,\infty$.

In the case of Gaussian states, all measures of entanglement are equivalent, and they are defined by the covariance matrix  
\begin{equation}
\boldsymbol{\sigma}=
\begin{pmatrix}
   \boldsymbol{\alpha} & \boldsymbol{\gamma}\\
   \boldsymbol{\gamma}^T & \boldsymbol{\beta}
\end{pmatrix}\,.
\end{equation}
Here, we define the matrices
\begin{equation}
   \boldsymbol{\alpha}\equiv
   \begin{pmatrix}
      \alpha_{1} & \alpha_{3}\\
      \alpha_{3} & \alpha_{2}
   \end{pmatrix}
   \!,\,
   \boldsymbol{\beta}\equiv
   \begin{pmatrix}
      \beta_{1} & \beta_{3}\\
      \beta_{3} & \beta_{2}
   \end{pmatrix}
   \!,\,
   \boldsymbol{\gamma}\equiv
   \begin{pmatrix}
      \gamma_{11} & \gamma_{12}\\
      \gamma_{21} & \gamma_{22}
   \end{pmatrix}
\end{equation}
with
\begin{align}
   \alpha_{1}&=
   \langle\hat s_1^2\rangle+\langle(\hat s_1^\dag)_{}^2\rangle
   +2\langle\hat s_1^\dag\hat s_1^{}\rangle
   -\langle\hat s_1^{}+\hat s_1^\dag\rangle_{}^2+1
   \\ 
   \alpha_{2}&=
   -\langle\hat s_1^2\rangle-\langle(\hat s_1^\dag)_{}^2\rangle
   +2\langle\hat s_1^\dag\hat s_1^{}\rangle
   +\langle\hat s_1^{}-\hat s_1^\dag\rangle_{}^2+1
   \\
   \alpha_{3}&=
   i\big(-\langle\hat s_1^2\rangle+\langle(\hat s_1^\dag)_{}^2\rangle
   +\langle\hat s_1^{}\rangle_{}^2-\langle\hat s_1^\dag\rangle_{}^2\big)
   \\
   \beta_{1}&=
   \langle\hat s_2^2\rangle+\langle(\hat s_2^\dag)_{}^2\rangle
   +2\langle\hat s_2^\dag\hat s_2^{}\rangle
   -\langle\hat s_2^{}+\hat s_2^\dag\rangle_{}^2+1
   \\ 
   \beta_{2}&=
   -\langle\hat s_2^2\rangle-\langle(\hat s_2^\dag)_{}^2\rangle
   +2\langle\hat s_2^\dag\hat s_2^{}\rangle
   +\langle\hat s_2^{}-\hat s_2^\dag\rangle_{}^2+1
   \\
   \beta_{3}&=
   i\big(-\langle\hat s_2^2\rangle+\langle(\hat s_2^\dag)_{}^2\rangle
   +\langle\hat s_2^{}\rangle_{}^2-\langle\hat s_2^\dag\rangle_{}^2\big)
   \\
   \gamma_{11}&=
   \langle\hat s_1^{}\hat s_2^{}+\hat s_1^{}\hat s_2^\dag
   +\hat s_1^\dag\hat s_2^{}+\hat s_1^\dag\hat s_2^\dag\rangle/2
   \nonumber\\&
   \quad+\langle\hat s_2^{}\hat s_1^{}+\hat s_2^{}\hat s_1^\dag
   +\hat s_2^\dag\hat s_1^{}+\hat s_2^\dag\hat s_1^\dag\rangle/2
   \nonumber\\&\quad
   -\langle\hat s_1^{}+\hat s_1^\dag\rangle
   \langle\hat s_2^{}+\hat s_2^\dag\rangle
   \\
   \gamma_{12}&=
   \langle\hat s_1^{}\hat s_2^{}-\hat s_1^{}\hat s_2^\dag
   +\hat s_1^\dag\hat s_2^{}-\hat s_1^\dag\hat s_2^\dag\rangle/2i
   \nonumber\\&\quad
   +\langle\hat s_2^{}\hat s_1^{}+\hat s_2^{}\hat s_1^\dag
   -\hat s_2^\dag\hat s_1^{}-\hat s_2^\dag\hat s_1^\dag\rangle/2i
   \nonumber\\&\quad
   +i\langle\hat s_1^{}+\hat s_1^\dag\rangle
   \langle\hat s_2^{}-\hat s_2^\dag\rangle
   \\
   \gamma_{21}&=
   \langle\hat s_1^{}\hat s_2^{}+\hat s_1^{}\hat s_2^\dag
   -\hat s_1^\dag\hat s_2^{}-\hat s_1^\dag\hat s_2^\dag\rangle/2i
   \nonumber\\&\quad
   +\langle\hat s_2^{}\hat s_1^{}-\hat s_2^{}\hat s_1^\dag
   +\hat s_2^\dag\hat s_1^{}-\hat s_2^\dag\hat s_1^\dag\rangle/2i
   \nonumber\\&\quad
   +i\langle\hat s_1^{}-\hat s_1^\dag\rangle
   \langle\hat s_2^{}+\hat s_2^\dag\rangle
   \\
   \gamma_{22}&=
   \langle-\hat s_1^{}\hat s_2^{}+\hat s_1^{}\hat s_2^\dag
   +\hat s_1^\dag\hat s_2^{}-\hat s_1^\dag\hat s_2^\dag\rangle/2
   \nonumber\\&\quad
   +\langle-\hat s_2^{}\hat s_1^{}+\hat s_2^{}\hat s_1^\dag
   +\hat s_2^\dag\hat s_1^{}-\hat s_2^\dag\hat s_1^\dag\rangle/2
   \nonumber\\&\quad
   +\langle\hat s_1^{}-\hat s_1^\dag\rangle
   \langle\hat s_2^{}-\hat s_2^\dag\rangle
\end{align}
and $\boldsymbol{\gamma}^T$ being the transpose of $\boldsymbol{\gamma}$.  Finally, the negativity becomes~\cite{Adesso:2005a}
\begin{equation}
   {\cal N}=\max\left\{0,\frac{1-\nu}{2\nu}\right\}\,\equiv \max\left\{0,\tilde{\mathcal{N}}\right\},
\end{equation}
where $\nu\,{\equiv}\,\sqrt{\left(\Delta(\boldsymbol{\sigma})-\sqrt{\Delta_{}^2(\boldsymbol{\sigma})-4\det\boldsymbol{\sigma}}\right)/2}$ and $\Delta(\boldsymbol{\sigma})\,{\equiv}\,\det\boldsymbol{\alpha}+\det\boldsymbol{\beta}-2\det\boldsymbol{\gamma}$.\\
Note that, despite not being a measure, the negativity kernel $\tilde{\mathcal{N}}$ is a witness for arbitrary bipartite entanglement. In fact, if a non-Gaussian state has the same first and second moments as an entangled Gaussian state, it is entangled~\cite{Hyllus:2006a}. Consequently, $\tilde{\mathcal{N}}\,{>}\,0$ implies entanglement for any bipartite state.

\subsection{Verifying consistency with a Gaussian state using higher order cumulants}
\label{sec:gaussianity}

In order to check whether the states we reconstruct are consistent with Gaussian states, we evaluate the $(l\,{+}\,m)^{\rm th}$ order cumulants $\langle\langle (\hat a_{}^l)^\dag \hat a^m\rangle\rangle$ for $l\,{+}\,m\,{\le}\,4$ and $l,m\,{\in}\,\mathbb{N}_0$. Equivalently to the moments, the cumulants describe a probability distribution. The definition of cumulant for a quantum state with density matrix $\rho$ can be written as~\cite{Schack:1990a}
\begin{align}
\label{cumulant}
&\langle\langle (\hat a_{}^l)_{}^\dag \hat a_{}^m\rangle\rangle \equiv \nonumber\\
&\frac{\partial_{}^l}{\partial(i\beta^*)_{}^l}\frac{\partial_{}^m}{\partial(i\beta)_{}^m}\left[ \frac{\beta\beta^* }{2}
+\ln {\rm Tr}\left(e^{(i\beta^*\hat a^\dag + i\beta \hat a)}\rho\right)  \right]_{\beta,\beta^*=0}\,.
\end{align}
Only Gaussian states have a finite number of nonzero cumulants. More specifically, all their cumulants vanish for $l\,{+}\,m\,{>}\,2$. In other words, finding a nonzero cumulant of $3^{\rm rd}$ or higher order implies that the state is not Gaussian. Despite not being a strict proof, the fact that the $3^{\rm rd}$ and $4^{\rm th}$ order cumulant are very small or vanish in an experimental reconstruction constitutes a reasonable indication that the reconstructed state is Gaussian.

As an example we spell out the $3^{\rm rd}$ order cumulants as functions of the moments at the beam splitter outputs reconstructed with the reference-state method. We find~\cite{Jiang:2010a}
\begin{align}
\langle\langle \hat s_{1,2}^3\rangle\rangle&=\langle \hat s_{1,2}^3\rangle-3\langle  \hat s_{1,2}^2\rangle\langle \hat s_{1,2}^{}\rangle+2\langle\hat s_{1,2}^{}\rangle^3 \label{C03}\\
\langle\langle \hat s_{1,2}^\dag \hat s_{1,2}^2\rangle\rangle&=\langle\hat s_{1,2}^\dag \hat s_{1,2}^2 \rangle-\langle \hat s_{1,2}^\dag \rangle\langle\hat s_{1,2}^2\rangle\nonumber\\
&\quad-2\langle\hat s_{1,2}^\dag \hat s_{1,2}^{}\rangle\langle\hat s_{1,2}^{}\rangle+2\langle\hat s_{1,2}^\dag\rangle\langle \hat s_{1,2}^{}\rangle_{}^2   \label{C12}
\,.
\end{align} 

\begin{acknowledgments} 
The authors thank Christopher Eichler for discussions. We acknowledge support from the Deutsche Forschungsgemeinschaft via the Sonderforschungsbereich~631, the German excellence initiative via the `Nanosystems Initiative Munich' (NIM), from the EU projects SOLID, CCQED and PROMISCE, from MEXT Kakenhi ``Quantum Cybernetics'', the JSPS through its FIRST Program, the Project for Developing Innovation Systems of MEXT, the NICT Commissioned Research, EPSRC EP/H050434/1, Basque Government IT472-10, and Spanish MICINN FIS2009-12773-C02-01.
\end{acknowledgments}

\end{cbunit}

\begin{thebibliography}{31}%
\makeatletter
\providecommand \@ifxundefined [1]{%
 \@ifx{#1\undefined}
}%
\providecommand \@ifnum [1]{%
 \ifnum #1\expandafter \@firstoftwo
 \else \expandafter \@secondoftwo
 \fi
}%
\providecommand \@ifx [1]{%
 \ifx #1\expandafter \@firstoftwo
 \else \expandafter \@secondoftwo
 \fi
}%
\providecommand \natexlab [1]{#1}%
\providecommand \enquote  [1]{``#1''}%
\providecommand \bibnamefont  [1]{#1}%
\providecommand \bibfnamefont [1]{#1}%
\providecommand \citenamefont [1]{#1}%
\providecommand \href@noop [0]{\@secondoftwo}%
\providecommand \href [0]{\begingroup \@sanitize@url \@href}%
\providecommand \@href[1]{\@@startlink{#1}\@@href}%
\providecommand \@@href[1]{\endgroup#1\@@endlink}%
\providecommand \@sanitize@url [0]{\catcode `\\12\catcode `\$12\catcode
  `\&12\catcode `\#12\catcode `\^12\catcode `\_12\catcode `\%12\relax}%
\providecommand \@@startlink[1]{}%
\providecommand \@@endlink[0]{}%
\providecommand \url  [0]{\begingroup\@sanitize@url \@url }%
\providecommand \@url [1]{\endgroup\@href {#1}{\urlprefix }}%
\providecommand \urlprefix  [0]{URL }%
\providecommand \Eprint [0]{\href }%
\providecommand \doibase [0]{http://dx.doi.org/}%
\providecommand \selectlanguage [0]{\@gobble}%
\providecommand \bibinfo  [0]{\@secondoftwo}%
\providecommand \bibfield  [0]{\@secondoftwo}%
\providecommand \translation [1]{[#1]}%
\providecommand \BibitemOpen [0]{}%
\providecommand \bibitemStop [0]{}%
\providecommand \bibitemNoStop [0]{.\EOS\space}%
\providecommand \EOS [0]{\spacefactor3000\relax}%
\providecommand \BibitemShut  [1]{\csname bibitem#1\endcsname}%
\let\auto@bib@innerbib\@empty
%</preamble>
\bibitem [{\citenamefont {Horodecki}\ \emph {et~al.}(2009)\citenamefont
  {Horodecki}, \citenamefont {Horodecki}, \citenamefont {Horodecki},\ and\
  \citenamefont {Horodecki}}]{Horodecki:2009a}%
  \BibitemOpen
  \bibfield  {author} {\bibinfo {author} {\bibfnamefont {R.}~\bibnamefont
  {Horodecki}}, \bibinfo {author} {\bibfnamefont {P.}~\bibnamefont
  {Horodecki}}, \bibinfo {author} {\bibfnamefont {M.}~\bibnamefont
  {Horodecki}}, \ and\ \bibinfo {author} {\bibfnamefont {K.}~\bibnamefont
  {Horodecki}},\ }\href {\doibase 10.1103/RevModPhys.81.865} {\bibfield
  {journal} {\bibinfo  {journal} {Rev. Mod. Phys.}\ }\textbf {\bibinfo {volume}
  {81}},\ \bibinfo {pages} {865} (\bibinfo {year} {2009})}\BibitemShut
  {NoStop}%
\bibitem [{\citenamefont {Born}(1971)}]{Born:1971a}%
  \BibitemOpen
  \bibfield  {author} {\bibinfo {author} {\bibfnamefont {M.}~\bibnamefont
  {Born}},\ }\href@noop {} {\emph {\bibinfo {title} {The Born-Einstein
  Letters}}}\ (\bibinfo  {publisher} {Walker and Company},\ \bibinfo {address}
  {New York},\ \bibinfo {year} {1971})\BibitemShut {NoStop}%
\bibitem [{\citenamefont {Einstein}\ \emph {et~al.}(1935)\citenamefont
  {Einstein}, \citenamefont {Podolsky},\ and\ \citenamefont
  {Rosen}}]{Einstein:1935a}%
  \BibitemOpen
  \bibfield  {author} {\bibinfo {author} {\bibfnamefont {A.}~\bibnamefont
  {Einstein}}, \bibinfo {author} {\bibfnamefont {B.}~\bibnamefont {Podolsky}},
  \ and\ \bibinfo {author} {\bibfnamefont {N.}~\bibnamefont {Rosen}},\ }\href
  {\doibase 10.1103/PhysRev.47.777} {\bibfield  {journal} {\bibinfo  {journal}
  {Phys. Rev.}\ }\textbf {\bibinfo {volume} {47}},\ \bibinfo {pages} {777}
  (\bibinfo {year} {1935})}\BibitemShut {NoStop}%
\bibitem [{\citenamefont {Raimond}\ \emph {et~al.}(2001)\citenamefont
  {Raimond}, \citenamefont {Brune},\ and\ \citenamefont
  {Haroche}}]{Raimond:2001a}%
  \BibitemOpen
  \bibfield  {author} {\bibinfo {author} {\bibfnamefont {J.}~\bibnamefont
  {Raimond}}, \bibinfo {author} {\bibfnamefont {M.}~\bibnamefont {Brune}}, \
  and\ \bibinfo {author} {\bibfnamefont {S.}~\bibnamefont {Haroche}},\ }\href
  {\doibase 10.1103/RevModPhys.73.565} {\bibfield  {journal} {\bibinfo
  {journal} {Rev. Mod. Phys.}\ }\textbf {\bibinfo {volume} {73}},\ \bibinfo
  {pages} {565} (\bibinfo {year} {2001})}\BibitemShut {NoStop}%
\bibitem [{\citenamefont {Braunstein}\ and\ \citenamefont {van
  Loock}(2005)}]{Braunstein:2005a}%
  \BibitemOpen
  \bibfield  {author} {\bibinfo {author} {\bibfnamefont {S.~L.}\ \bibnamefont
  {Braunstein}}\ and\ \bibinfo {author} {\bibfnamefont {P.}~\bibnamefont {van
  Loock}},\ }\href {\doibase 10.1103/RevModPhys.77.513} {\bibfield  {journal}
  {\bibinfo  {journal} {Rev. Mod. Phys.}\ }\textbf {\bibinfo {volume} {77}},\
  \bibinfo {pages} {513} (\bibinfo {year} {2005})}\BibitemShut {NoStop}%
\bibitem [{\citenamefont {Ou}\ \emph {et~al.}(1992)\citenamefont {Ou},
  \citenamefont {Pereira}, \citenamefont {Kimble},\ and\ \citenamefont
  {Peng}}]{Ou:1992a}%
  \BibitemOpen
  \bibfield  {author} {\bibinfo {author} {\bibfnamefont {Z.~Y.}\ \bibnamefont
  {Ou}}, \bibinfo {author} {\bibfnamefont {S.~F.}\ \bibnamefont {Pereira}},
  \bibinfo {author} {\bibfnamefont {H.~J.}\ \bibnamefont {Kimble}}, \ and\
  \bibinfo {author} {\bibfnamefont {K.~C.}\ \bibnamefont {Peng}},\ }\href
  {\doibase 10.1103/PhysRevLett.68.3663} {\bibfield  {journal} {\bibinfo
  {journal} {Phys. Rev. Lett.}\ }\textbf {\bibinfo {volume} {68}},\ \bibinfo
  {pages} {3663} (\bibinfo {year} {1992})}\BibitemShut {NoStop}%
\bibitem [{\citenamefont {Schoelkopf}\ and\ \citenamefont
  {Girvin}(2008)}]{Schoelkopf:2008a}%
  \BibitemOpen
  \bibfield  {author} {\bibinfo {author} {\bibfnamefont {R.~J.}\ \bibnamefont
  {Schoelkopf}}\ and\ \bibinfo {author} {\bibfnamefont {S.~M.}\ \bibnamefont
  {Girvin}},\ }\href@noop {} {\bibfield  {journal} {\bibinfo  {journal}
  {Nature}\ }\textbf {\bibinfo {volume} {451}},\ \bibinfo {pages} {664}
  (\bibinfo {year} {2008})}\BibitemShut {NoStop}%
\bibitem [{\citenamefont {Clarke}\ and\ \citenamefont
  {Wilhelm}(2008)}]{Clarke:2008a}%
  \BibitemOpen
  \bibfield  {author} {\bibinfo {author} {\bibfnamefont {J.}~\bibnamefont
  {Clarke}}\ and\ \bibinfo {author} {\bibfnamefont {F.~K.}\ \bibnamefont
  {Wilhelm}},\ }\href@noop {} {\bibfield  {journal} {\bibinfo  {journal}
  {Nature}\ }\textbf {\bibinfo {volume} {453}},\ \bibinfo {pages} {1031}
  (\bibinfo {year} {2008})}\BibitemShut {NoStop}%
\bibitem [{\citenamefont {Mariantoni}\ \emph {et~al.}(2011)\citenamefont
  {Mariantoni}, \citenamefont {Wang}, \citenamefont {Yamamoto}, \citenamefont
  {Neeley}, \citenamefont {Bialczak}, \citenamefont {Chen}, \citenamefont
  {Lenander}, \citenamefont {Lucero}, \citenamefont {{O'Connell}},
  \citenamefont {Sank}, \citenamefont {Weides}, \citenamefont {Wenner},
  \citenamefont {Yin}, \citenamefont {Zhao}, \citenamefont {Korotkov},
  \citenamefont {Cleland},\ and\ \citenamefont {Martinis}}]{Mariantoni:2011a}%
  \BibitemOpen
  \bibfield  {author} {\bibinfo {author} {\bibfnamefont {M.}~\bibnamefont
  {Mariantoni}}, \bibinfo {author} {\bibfnamefont {H.}~\bibnamefont {Wang}},
  \bibinfo {author} {\bibfnamefont {T.}~\bibnamefont {Yamamoto}}, \bibinfo
  {author} {\bibfnamefont {M.}~\bibnamefont {Neeley}}, \bibinfo {author}
  {\bibfnamefont {R.~C.}\ \bibnamefont {Bialczak}}, \bibinfo {author}
  {\bibfnamefont {Y.}~\bibnamefont {Chen}}, \bibinfo {author} {\bibfnamefont
  {M.}~\bibnamefont {Lenander}}, \bibinfo {author} {\bibfnamefont
  {E.}~\bibnamefont {Lucero}}, \bibinfo {author} {\bibfnamefont {A.~D.}\
  \bibnamefont {{O'Connell}}}, \bibinfo {author} {\bibfnamefont
  {D.}~\bibnamefont {Sank}}, \bibinfo {author} {\bibfnamefont {M.}~\bibnamefont
  {Weides}}, \bibinfo {author} {\bibfnamefont {J.}~\bibnamefont {Wenner}},
  \bibinfo {author} {\bibfnamefont {Y.}~\bibnamefont {Yin}}, \bibinfo {author}
  {\bibfnamefont {J.}~\bibnamefont {Zhao}}, \bibinfo {author} {\bibfnamefont
  {A.~N.}\ \bibnamefont {Korotkov}}, \bibinfo {author} {\bibfnamefont {A.~N.}\
  \bibnamefont {Cleland}}, \ and\ \bibinfo {author} {\bibfnamefont {J.~M.}\
  \bibnamefont {Martinis}},\ }\href {\doibase 10.1126/science.1208517}
  {\bibfield  {journal} {\bibinfo  {journal} {Science}\ }\textbf {\bibinfo
  {volume} {334}},\ \bibinfo {pages} {61} (\bibinfo {year} {2011})}\BibitemShut
  {NoStop}%
\bibitem [{\citenamefont {Wallraff}\ \emph {et~al.}(2004)\citenamefont
  {Wallraff}, \citenamefont {Schuster}, \citenamefont {Blais}, \citenamefont
  {Frunzio}, \citenamefont {Huang}, \citenamefont {Majer}, \citenamefont
  {Kumar}, \citenamefont {Girvin},\ and\ \citenamefont
  {Schoelkopf}}]{Wallraff:2004a}%
  \BibitemOpen
  \bibfield  {author} {\bibinfo {author} {\bibfnamefont {A.}~\bibnamefont
  {Wallraff}}, \bibinfo {author} {\bibfnamefont {D.~I.}\ \bibnamefont
  {Schuster}}, \bibinfo {author} {\bibfnamefont {A.}~\bibnamefont {Blais}},
  \bibinfo {author} {\bibfnamefont {L.}~\bibnamefont {Frunzio}}, \bibinfo
  {author} {\bibfnamefont {R.-S.}\ \bibnamefont {Huang}}, \bibinfo {author}
  {\bibfnamefont {J.}~\bibnamefont {Majer}}, \bibinfo {author} {\bibfnamefont
  {S.}~\bibnamefont {Kumar}}, \bibinfo {author} {\bibfnamefont {S.~M.}\
  \bibnamefont {Girvin}}, \ and\ \bibinfo {author} {\bibfnamefont {R.~J.}\
  \bibnamefont {Schoelkopf}},\ }\href@noop {} {\bibfield  {journal} {\bibinfo
  {journal} {Nature}\ }\textbf {\bibinfo {volume} {431}},\ \bibinfo {pages}
  {162} (\bibinfo {year} {2004})}\BibitemShut {NoStop}%
\bibitem [{\citenamefont {Niemczyk}\ \emph {et~al.}(2010)\citenamefont
  {Niemczyk}, \citenamefont {Deppe}, \citenamefont {Huebl}, \citenamefont
  {Menzel}, \citenamefont {Hocke}, \citenamefont {Schwarz}, \citenamefont
  {{Garcia-Ripoll}}, \citenamefont {Zueco}, \citenamefont {H{\"u}mmer},
  \citenamefont {Solano}, \citenamefont {Marx},\ and\ \citenamefont
  {Gross}}]{Niemczyk:2010a}%
  \BibitemOpen
  \bibfield  {author} {\bibinfo {author} {\bibfnamefont {T.}~\bibnamefont
  {Niemczyk}}, \bibinfo {author} {\bibfnamefont {F.}~\bibnamefont {Deppe}},
  \bibinfo {author} {\bibfnamefont {H.}~\bibnamefont {Huebl}}, \bibinfo
  {author} {\bibfnamefont {E.~P.}\ \bibnamefont {Menzel}}, \bibinfo {author}
  {\bibfnamefont {F.}~\bibnamefont {Hocke}}, \bibinfo {author} {\bibfnamefont
  {M.~J.}\ \bibnamefont {Schwarz}}, \bibinfo {author} {\bibfnamefont {J.~J.}\
  \bibnamefont {{Garcia-Ripoll}}}, \bibinfo {author} {\bibfnamefont
  {D.}~\bibnamefont {Zueco}}, \bibinfo {author} {\bibfnamefont
  {T.}~\bibnamefont {H{\"u}mmer}}, \bibinfo {author} {\bibfnamefont
  {E.}~\bibnamefont {Solano}}, \bibinfo {author} {\bibfnamefont
  {A.}~\bibnamefont {Marx}}, \ and\ \bibinfo {author} {\bibfnamefont
  {R.}~\bibnamefont {Gross}},\ }\href {\doibase 10.1038/nphys1730} {\bibfield
  {journal} {\bibinfo  {journal} {Nature Phys.}\ }\textbf {\bibinfo {volume}
  {6}},\ \bibinfo {pages} {772} (\bibinfo {year} {2010})}\BibitemShut {NoStop}%
\bibitem [{\citenamefont {Majer}\ \emph {et~al.}(2007)\citenamefont {Majer},
  \citenamefont {Chow}, \citenamefont {Gambetta}, \citenamefont {Koch},
  \citenamefont {Johnson}, \citenamefont {Schreier}, \citenamefont {Frunzio},
  \citenamefont {Schuster}, \citenamefont {Houck}, \citenamefont {Wallraff},
  \citenamefont {Blais}, \citenamefont {Devoret}, \citenamefont {Girvin},\ and\
  \citenamefont {Schoelkopf}}]{Majer:2007a}%
  \BibitemOpen
  \bibfield  {author} {\bibinfo {author} {\bibfnamefont {J.}~\bibnamefont
  {Majer}}, \bibinfo {author} {\bibfnamefont {J.~M.}\ \bibnamefont {Chow}},
  \bibinfo {author} {\bibfnamefont {J.~M.}\ \bibnamefont {Gambetta}}, \bibinfo
  {author} {\bibfnamefont {J.}~\bibnamefont {Koch}}, \bibinfo {author}
  {\bibfnamefont {B.~R.}\ \bibnamefont {Johnson}}, \bibinfo {author}
  {\bibfnamefont {J.~A.}\ \bibnamefont {Schreier}}, \bibinfo {author}
  {\bibfnamefont {L.}~\bibnamefont {Frunzio}}, \bibinfo {author} {\bibfnamefont
  {D.~I.}\ \bibnamefont {Schuster}}, \bibinfo {author} {\bibfnamefont {A.~A.}\
  \bibnamefont {Houck}}, \bibinfo {author} {\bibfnamefont {A.}~\bibnamefont
  {Wallraff}}, \bibinfo {author} {\bibfnamefont {A.}~\bibnamefont {Blais}},
  \bibinfo {author} {\bibfnamefont {M.~H.}\ \bibnamefont {Devoret}}, \bibinfo
  {author} {\bibfnamefont {S.~M.}\ \bibnamefont {Girvin}}, \ and\ \bibinfo
  {author} {\bibfnamefont {R.~J.}\ \bibnamefont {Schoelkopf}},\ }\href@noop {}
  {\bibfield  {journal} {\bibinfo  {journal} {Nature}\ }\textbf {\bibinfo
  {volume} {449}},\ \bibinfo {pages} {443} (\bibinfo {year}
  {2007})}\BibitemShut {NoStop}%
\bibitem [{\citenamefont {Sillanp{\"a}{\"a}}\ \emph {et~al.}(2007)\citenamefont
  {Sillanp{\"a}{\"a}}, \citenamefont {Park},\ and\ \citenamefont
  {Simmonds}}]{Sillanpaa:2007a}%
  \BibitemOpen
  \bibfield  {author} {\bibinfo {author} {\bibfnamefont {M.~A.}\ \bibnamefont
  {Sillanp{\"a}{\"a}}}, \bibinfo {author} {\bibfnamefont {J.~I.}\ \bibnamefont
  {Park}}, \ and\ \bibinfo {author} {\bibfnamefont {R.~W.}\ \bibnamefont
  {Simmonds}},\ }\href@noop {} {\bibfield  {journal} {\bibinfo  {journal}
  {Nature}\ }\textbf {\bibinfo {volume} {449}},\ \bibinfo {pages} {438}
  (\bibinfo {year} {2007})}\BibitemShut {NoStop}%
\bibitem [{\citenamefont {Yamamoto}\ \emph {et~al.}(2003)\citenamefont
  {Yamamoto}, \citenamefont {Pashkin}, \citenamefont {Astafiev}, \citenamefont
  {Nakamura},\ and\ \citenamefont {Tsai}}]{Yamamoto:2003a}%
  \BibitemOpen
  \bibfield  {author} {\bibinfo {author} {\bibfnamefont {T.}~\bibnamefont
  {Yamamoto}}, \bibinfo {author} {\bibfnamefont {{\relax Yu}.~A.}\ \bibnamefont
  {Pashkin}}, \bibinfo {author} {\bibfnamefont {O.}~\bibnamefont {Astafiev}},
  \bibinfo {author} {\bibfnamefont {Y.}~\bibnamefont {Nakamura}}, \ and\
  \bibinfo {author} {\bibfnamefont {J.~S.}\ \bibnamefont {Tsai}},\ }\href
  {\doibase 10.1038/nature02015} {\bibfield  {journal} {\bibinfo  {journal}
  {Nature}\ }\textbf {\bibinfo {volume} {425}},\ \bibinfo {pages} {941}
  (\bibinfo {year} {2003})}\BibitemShut {NoStop}%
\bibitem [{\citenamefont {Plantenberg}\ \emph {et~al.}(2007)\citenamefont
  {Plantenberg}, \citenamefont {de~Groot}, \citenamefont {Harmans},\ and\
  \citenamefont {Mooij}}]{Plantenberg:2007a}%
  \BibitemOpen
  \bibfield  {author} {\bibinfo {author} {\bibfnamefont {J.~H.}\ \bibnamefont
  {Plantenberg}}, \bibinfo {author} {\bibfnamefont {P.~C.}\ \bibnamefont
  {de~Groot}}, \bibinfo {author} {\bibfnamefont {C.~J. P.~M.}\ \bibnamefont
  {Harmans}}, \ and\ \bibinfo {author} {\bibfnamefont {J.~E.}\ \bibnamefont
  {Mooij}},\ }\href@noop {} {\bibfield  {journal} {\bibinfo  {journal}
  {Nature}\ }\textbf {\bibinfo {volume} {447}},\ \bibinfo {pages} {836}
  (\bibinfo {year} {2007})}\BibitemShut {NoStop}%
\bibitem [{\citenamefont {Ansmann}\ \emph {et~al.}(2009)\citenamefont
  {Ansmann}, \citenamefont {Wang}, \citenamefont {Bialczak}, \citenamefont
  {Hofheinz}, \citenamefont {Lucero}, \citenamefont {Neeley}, \citenamefont
  {{O'Connell}}, \citenamefont {Sank}, \citenamefont {Weides}, \citenamefont
  {Wenner}, \citenamefont {Cleland},\ and\ \citenamefont
  {Martinis}}]{Ansmann:2009a}%
  \BibitemOpen
  \bibfield  {author} {\bibinfo {author} {\bibfnamefont {M.}~\bibnamefont
  {Ansmann}}, \bibinfo {author} {\bibfnamefont {H.}~\bibnamefont {Wang}},
  \bibinfo {author} {\bibfnamefont {R.~C.}\ \bibnamefont {Bialczak}}, \bibinfo
  {author} {\bibfnamefont {M.}~\bibnamefont {Hofheinz}}, \bibinfo {author}
  {\bibfnamefont {E.}~\bibnamefont {Lucero}}, \bibinfo {author} {\bibfnamefont
  {M.}~\bibnamefont {Neeley}}, \bibinfo {author} {\bibfnamefont {A.~D.}\
  \bibnamefont {{O'Connell}}}, \bibinfo {author} {\bibfnamefont
  {D.}~\bibnamefont {Sank}}, \bibinfo {author} {\bibfnamefont {M.}~\bibnamefont
  {Weides}}, \bibinfo {author} {\bibfnamefont {J.}~\bibnamefont {Wenner}},
  \bibinfo {author} {\bibfnamefont {A.~N.}\ \bibnamefont {Cleland}}, \ and\
  \bibinfo {author} {\bibfnamefont {J.~M.}\ \bibnamefont {Martinis}},\ }\href
  {\doibase 10.1038/nature08363} {\bibfield  {journal} {\bibinfo  {journal}
  {Nature}\ }\textbf {\bibinfo {volume} {461}},\ \bibinfo {pages} {504}
  (\bibinfo {year} {2009})}\BibitemShut {NoStop}%
\bibitem [{\citenamefont {Mariantoni}\ \emph {et~al.}(2010)\citenamefont
  {Mariantoni}, \citenamefont {Menzel}, \citenamefont {Deppe}, \citenamefont
  {{Araque Caballero}}, \citenamefont {Baust}, \citenamefont {Niemczyk},
  \citenamefont {Hoffmann}, \citenamefont {Solano}, \citenamefont {Marx},\ and\
  \citenamefont {Gross}}]{Mariantoni:2010a}%
  \BibitemOpen
  \bibfield  {author} {\bibinfo {author} {\bibfnamefont {M.}~\bibnamefont
  {Mariantoni}}, \bibinfo {author} {\bibfnamefont {E.~P.}\ \bibnamefont
  {Menzel}}, \bibinfo {author} {\bibfnamefont {F.}~\bibnamefont {Deppe}},
  \bibinfo {author} {\bibfnamefont {M.~{\'A}.}\ \bibnamefont {{Araque
  Caballero}}}, \bibinfo {author} {\bibfnamefont {A.}~\bibnamefont {Baust}},
  \bibinfo {author} {\bibfnamefont {T.}~\bibnamefont {Niemczyk}}, \bibinfo
  {author} {\bibfnamefont {E.}~\bibnamefont {Hoffmann}}, \bibinfo {author}
  {\bibfnamefont {E.}~\bibnamefont {Solano}}, \bibinfo {author} {\bibfnamefont
  {A.}~\bibnamefont {Marx}}, \ and\ \bibinfo {author} {\bibfnamefont
  {R.}~\bibnamefont {Gross}},\ }\href {\doibase 10.1103/PhysRevLett.105.133601}
  {\bibfield  {journal} {\bibinfo  {journal} {Phys. Rev. Lett.}\ }\textbf
  {\bibinfo {volume} {105}},\ \bibinfo {pages} {133601} (\bibinfo {year}
  {2010})}\BibitemShut {NoStop}%
\bibitem [{\citenamefont {Menzel}\ \emph {et~al.}(2010)\citenamefont {Menzel},
  \citenamefont {Deppe}, \citenamefont {Mariantoni}, \citenamefont {{Araque
  Caballero}}, \citenamefont {Baust}, \citenamefont {Niemczyk}, \citenamefont
  {Hoffmann}, \citenamefont {Marx}, \citenamefont {Solano},\ and\ \citenamefont
  {Gross}}]{Menzel:2010a}%
  \BibitemOpen
  \bibfield  {author} {\bibinfo {author} {\bibfnamefont {E.~P.}\ \bibnamefont
  {Menzel}}, \bibinfo {author} {\bibfnamefont {F.}~\bibnamefont {Deppe}},
  \bibinfo {author} {\bibfnamefont {M.}~\bibnamefont {Mariantoni}}, \bibinfo
  {author} {\bibfnamefont {M.~{\'A}.}\ \bibnamefont {{Araque Caballero}}},
  \bibinfo {author} {\bibfnamefont {A.}~\bibnamefont {Baust}}, \bibinfo
  {author} {\bibfnamefont {T.}~\bibnamefont {Niemczyk}}, \bibinfo {author}
  {\bibfnamefont {E.}~\bibnamefont {Hoffmann}}, \bibinfo {author}
  {\bibfnamefont {A.}~\bibnamefont {Marx}}, \bibinfo {author} {\bibfnamefont
  {E.}~\bibnamefont {Solano}}, \ and\ \bibinfo {author} {\bibfnamefont
  {R.}~\bibnamefont {Gross}},\ }\href {\doibase 10.1103/PhysRevLett.105.100401}
  {\bibfield  {journal} {\bibinfo  {journal} {Phys. Rev. Lett.}\ }\textbf
  {\bibinfo {volume} {105}},\ \bibinfo {pages} {100401} (\bibinfo {year}
  {2010})}\BibitemShut {NoStop}%
\bibitem [{\citenamefont {Bozyigit}\ \emph {et~al.}(2011)\citenamefont
  {Bozyigit}, \citenamefont {Lang}, \citenamefont {Steffen}, \citenamefont
  {Fink}, \citenamefont {Eichler}, \citenamefont {Baur}, \citenamefont
  {Bianchetti}, \citenamefont {Leek}, \citenamefont {Filipp}, \citenamefont
  {da~Silva}, \citenamefont {Blais},\ and\ \citenamefont
  {Wallraff}}]{Bozyigit:2011a}%
  \BibitemOpen
  \bibfield  {author} {\bibinfo {author} {\bibfnamefont {D.}~\bibnamefont
  {Bozyigit}}, \bibinfo {author} {\bibfnamefont {C.}~\bibnamefont {Lang}},
  \bibinfo {author} {\bibfnamefont {L.}~\bibnamefont {Steffen}}, \bibinfo
  {author} {\bibfnamefont {J.~M.}\ \bibnamefont {Fink}}, \bibinfo {author}
  {\bibfnamefont {C.}~\bibnamefont {Eichler}}, \bibinfo {author} {\bibfnamefont
  {M.}~\bibnamefont {Baur}}, \bibinfo {author} {\bibfnamefont {R.}~\bibnamefont
  {Bianchetti}}, \bibinfo {author} {\bibfnamefont {P.~J.}\ \bibnamefont
  {Leek}}, \bibinfo {author} {\bibfnamefont {S.}~\bibnamefont {Filipp}},
  \bibinfo {author} {\bibfnamefont {M.~P.}\ \bibnamefont {da~Silva}}, \bibinfo
  {author} {\bibfnamefont {A.}~\bibnamefont {Blais}}, \ and\ \bibinfo {author}
  {\bibfnamefont {A.}~\bibnamefont {Wallraff}},\ }\href {\doibase
  10.1038/nphys1845} {\bibfield  {journal} {\bibinfo  {journal} {Nature Phys.}\
  }\textbf {\bibinfo {volume} {7}},\ \bibinfo {pages} {154} (\bibinfo {year}
  {2011})}\BibitemShut {NoStop}%
\bibitem [{\citenamefont {Mallet}\ \emph {et~al.}(2011)\citenamefont {Mallet},
  \citenamefont {Castellanos-Beltran}, \citenamefont {Ku}, \citenamefont
  {Glancy}, \citenamefont {Knill}, \citenamefont {Irwin}, \citenamefont
  {Hilton}, \citenamefont {Vale},\ and\ \citenamefont
  {Lehnert}}]{Mallet:2011a}%
  \BibitemOpen
  \bibfield  {author} {\bibinfo {author} {\bibfnamefont {F.}~\bibnamefont
  {Mallet}}, \bibinfo {author} {\bibfnamefont {M.~A.}\ \bibnamefont
  {Castellanos-Beltran}}, \bibinfo {author} {\bibfnamefont {H.~S.}\
  \bibnamefont {Ku}}, \bibinfo {author} {\bibfnamefont {S.}~\bibnamefont
  {Glancy}}, \bibinfo {author} {\bibfnamefont {E.}~\bibnamefont {Knill}},
  \bibinfo {author} {\bibfnamefont {K.~D.}\ \bibnamefont {Irwin}}, \bibinfo
  {author} {\bibfnamefont {G.~C.}\ \bibnamefont {Hilton}}, \bibinfo {author}
  {\bibfnamefont {L.~R.}\ \bibnamefont {Vale}}, \ and\ \bibinfo {author}
  {\bibfnamefont {K.~W.}\ \bibnamefont {Lehnert}},\ }\href {\doibase
  10.1103/PhysRevLett.106.220502} {\bibfield  {journal} {\bibinfo  {journal}
  {Phys. Rev. Lett.}\ }\textbf {\bibinfo {volume} {106}},\ \bibinfo {pages}
  {220502} (\bibinfo {year} {2011})}\BibitemShut {NoStop}%
\bibitem [{\citenamefont {Eichler}\ \emph
  {et~al.}(2011{\natexlab{a}})\citenamefont {Eichler}, \citenamefont
  {Bozyigit}, \citenamefont {Lang}, \citenamefont {Baur}, \citenamefont
  {Steffen}, \citenamefont {Fink}, \citenamefont {Filipp},\ and\ \citenamefont
  {Wallraff}}]{Eichler:2011b}%
  \BibitemOpen
  \bibfield  {author} {\bibinfo {author} {\bibfnamefont {C.}~\bibnamefont
  {Eichler}}, \bibinfo {author} {\bibfnamefont {D.}~\bibnamefont {Bozyigit}},
  \bibinfo {author} {\bibfnamefont {C.}~\bibnamefont {Lang}}, \bibinfo {author}
  {\bibfnamefont {M.}~\bibnamefont {Baur}}, \bibinfo {author} {\bibfnamefont
  {L.}~\bibnamefont {Steffen}}, \bibinfo {author} {\bibfnamefont {J.~M.}\
  \bibnamefont {Fink}}, \bibinfo {author} {\bibfnamefont {S.}~\bibnamefont
  {Filipp}}, \ and\ \bibinfo {author} {\bibfnamefont {A.}~\bibnamefont
  {Wallraff}},\ }\href {\doibase 10.1103/PhysRevLett.107.113601} {\bibfield
  {journal} {\bibinfo  {journal} {Phys. Rev. Lett.}\ }\textbf {\bibinfo
  {volume} {107}},\ \bibinfo {pages} {113601} (\bibinfo {year}
  {2011}{\natexlab{a}})}\BibitemShut {NoStop}%
\bibitem [{\citenamefont {Wilson}\ \emph {et~al.}(2011)\citenamefont {Wilson},
  \citenamefont {Johansson}, \citenamefont {Pourkabirian}, \citenamefont
  {Simoen}, \citenamefont {Johansson}, \citenamefont {Duty}, \citenamefont
  {Nori},\ and\ \citenamefont {Delsing}}]{Wilson:2011a}%
  \BibitemOpen
  \bibfield  {author} {\bibinfo {author} {\bibfnamefont {C.~M.}\ \bibnamefont
  {Wilson}}, \bibinfo {author} {\bibfnamefont {G.}~\bibnamefont {Johansson}},
  \bibinfo {author} {\bibfnamefont {A.}~\bibnamefont {Pourkabirian}}, \bibinfo
  {author} {\bibfnamefont {M.}~\bibnamefont {Simoen}}, \bibinfo {author}
  {\bibfnamefont {J.~R.}\ \bibnamefont {Johansson}}, \bibinfo {author}
  {\bibfnamefont {T.}~\bibnamefont {Duty}}, \bibinfo {author} {\bibfnamefont
  {F.}~\bibnamefont {Nori}}, \ and\ \bibinfo {author} {\bibfnamefont
  {P.}~\bibnamefont {Delsing}},\ }\href {\doibase 10.1038/nature10561}
  {\bibfield  {journal} {\bibinfo  {journal} {Nature}\ }\textbf {\bibinfo
  {volume} {479}},\ \bibinfo {pages} {376} (\bibinfo {year}
  {2011})}\BibitemShut {NoStop}%
\bibitem [{\citenamefont {Flurin}\ \emph {et~al.}(2012)\citenamefont {Flurin},
  \citenamefont {Roch}, \citenamefont {Mallet}, \citenamefont {Devoret},\ and\
  \citenamefont {Huard}}]{Flurin:2012a}%
  \BibitemOpen
  \bibfield  {author} {\bibinfo {author} {\bibfnamefont {E.}~\bibnamefont
  {Flurin}}, \bibinfo {author} {\bibfnamefont {N.}~\bibnamefont {Roch}},
  \bibinfo {author} {\bibfnamefont {F.}~\bibnamefont {Mallet}}, \bibinfo
  {author} {\bibfnamefont {M.~H.}\ \bibnamefont {Devoret}}, \ and\ \bibinfo
  {author} {\bibfnamefont {B.}~\bibnamefont {Huard}},\ }\href@noop {} {}
  (\bibinfo {year} {2012}),\ \Eprint {http://arxiv.org/abs/arXiv:1204.0732}
  {arXiv:1204.0732} \BibitemShut {NoStop}%
\bibitem [{\citenamefont {Hoffmann}\ \emph {et~al.}(2010)\citenamefont
  {Hoffmann}, \citenamefont {Deppe}, \citenamefont {Niemczyk}, \citenamefont
  {Wirth}, \citenamefont {Menzel}, \citenamefont {Wild}, \citenamefont {Huebl},
  \citenamefont {Mariantoni}, \citenamefont {Wei{\ss}l}, \citenamefont
  {Lukashenko}, \citenamefont {Zhuravel}, \citenamefont {Ustinov},
  \citenamefont {Marx},\ and\ \citenamefont {Gross}}]{Hoffmann:2010a}%
  \BibitemOpen
  \bibfield  {author} {\bibinfo {author} {\bibfnamefont {E.}~\bibnamefont
  {Hoffmann}}, \bibinfo {author} {\bibfnamefont {F.}~\bibnamefont {Deppe}},
  \bibinfo {author} {\bibfnamefont {T.}~\bibnamefont {Niemczyk}}, \bibinfo
  {author} {\bibfnamefont {T.}~\bibnamefont {Wirth}}, \bibinfo {author}
  {\bibfnamefont {E.~P.}\ \bibnamefont {Menzel}}, \bibinfo {author}
  {\bibfnamefont {G.}~\bibnamefont {Wild}}, \bibinfo {author} {\bibfnamefont
  {H.}~\bibnamefont {Huebl}}, \bibinfo {author} {\bibfnamefont
  {M.}~\bibnamefont {Mariantoni}}, \bibinfo {author} {\bibfnamefont
  {T.}~\bibnamefont {Wei{\ss}l}}, \bibinfo {author} {\bibfnamefont
  {A.}~\bibnamefont {Lukashenko}}, \bibinfo {author} {\bibfnamefont {A.~P.}\
  \bibnamefont {Zhuravel}}, \bibinfo {author} {\bibfnamefont {A.~V.}\
  \bibnamefont {Ustinov}}, \bibinfo {author} {\bibfnamefont {A.}~\bibnamefont
  {Marx}}, \ and\ \bibinfo {author} {\bibfnamefont {R.}~\bibnamefont {Gross}},\
  }\href {\doibase 10.1063/1.3522650} {\bibfield  {journal} {\bibinfo
  {journal} {Appl. Phys. Lett.}\ }\textbf {\bibinfo {volume} {97}},\ \bibinfo
  {pages} {222508} (\bibinfo {year} {2010})}\BibitemShut {NoStop}%
\bibitem [{\citenamefont {Eichler}\ \emph
  {et~al.}(2011{\natexlab{b}})\citenamefont {Eichler}, \citenamefont
  {Bozyigit}, \citenamefont {Lang}, \citenamefont {Steffen}, \citenamefont
  {Fink},\ and\ \citenamefont {Wallraff}}]{Eichler:2011a}%
  \BibitemOpen
  \bibfield  {author} {\bibinfo {author} {\bibfnamefont {C.}~\bibnamefont
  {Eichler}}, \bibinfo {author} {\bibfnamefont {D.}~\bibnamefont {Bozyigit}},
  \bibinfo {author} {\bibfnamefont {C.}~\bibnamefont {Lang}}, \bibinfo {author}
  {\bibfnamefont {L.}~\bibnamefont {Steffen}}, \bibinfo {author} {\bibfnamefont
  {J.}~\bibnamefont {Fink}}, \ and\ \bibinfo {author} {\bibfnamefont
  {A.}~\bibnamefont {Wallraff}},\ }\href {\doibase
  10.1103/PhysRevLett.106.220503} {\bibfield  {journal} {\bibinfo  {journal}
  {Phys. Rev. Lett.}\ }\textbf {\bibinfo {volume} {106}},\ \bibinfo {pages}
  {220503} (\bibinfo {year} {2011}{\natexlab{b}})}\BibitemShut {NoStop}%
\bibitem [{\citenamefont {Shchukin}\ and\ \citenamefont
  {Vogel}(2005)}]{Shchukin:2005a}%
  \BibitemOpen
  \bibfield  {author} {\bibinfo {author} {\bibfnamefont {E.}~\bibnamefont
  {Shchukin}}\ and\ \bibinfo {author} {\bibfnamefont {W.}~\bibnamefont
  {Vogel}},\ }\href {\doibase 10.1103/PhysRevLett.95.230502} {\bibfield
  {journal} {\bibinfo  {journal} {Phys. Rev. Lett.}\ }\textbf {\bibinfo
  {volume} {95}},\ \bibinfo {pages} {230502} (\bibinfo {year}
  {2005})}\BibitemShut {NoStop}%
\bibitem [{\citenamefont {Adesso}\ and\ \citenamefont
  {Illuminati}(2005)}]{Adesso:2005a}%
  \BibitemOpen
  \bibfield  {author} {\bibinfo {author} {\bibfnamefont {G.}~\bibnamefont
  {Adesso}}\ and\ \bibinfo {author} {\bibfnamefont {F.}~\bibnamefont
  {Illuminati}},\ }\href {\doibase 10.1103/PhysRevA.72.032334} {\bibfield
  {journal} {\bibinfo  {journal} {Phys. Rev. A}\ }\textbf {\bibinfo {volume}
  {72}},\ \bibinfo {pages} {032334} (\bibinfo {year} {2005})}\BibitemShut
  {NoStop}%
\bibitem [{\citenamefont {Yamamoto}\ \emph {et~al.}(2008)\citenamefont
  {Yamamoto}, \citenamefont {Inomata}, \citenamefont {Watanabe}, \citenamefont
  {Matsuba}, \citenamefont {Miyazaki}, \citenamefont {Oliver}, \citenamefont
  {Nakamura},\ and\ \citenamefont {Tsai}}]{Yamamoto:2008a}%
  \BibitemOpen
  \bibfield  {author} {\bibinfo {author} {\bibfnamefont {T.}~\bibnamefont
  {Yamamoto}}, \bibinfo {author} {\bibfnamefont {K.}~\bibnamefont {Inomata}},
  \bibinfo {author} {\bibfnamefont {M.}~\bibnamefont {Watanabe}}, \bibinfo
  {author} {\bibfnamefont {K.}~\bibnamefont {Matsuba}}, \bibinfo {author}
  {\bibfnamefont {T.}~\bibnamefont {Miyazaki}}, \bibinfo {author}
  {\bibfnamefont {W.~D.}\ \bibnamefont {Oliver}}, \bibinfo {author}
  {\bibfnamefont {Y.}~\bibnamefont {Nakamura}}, \ and\ \bibinfo {author}
  {\bibfnamefont {J.~S.}\ \bibnamefont {Tsai}},\ }\href {\doibase
  10.1063/1.2964182} {\bibfield  {journal} {\bibinfo  {journal} {Appl. Phys.
  Lett.}\ }\textbf {\bibinfo {volume} {93}},\ \bibinfo {pages} {042510}
  (\bibinfo {year} {2008})}\BibitemShut {NoStop}%
\bibitem [{\citenamefont {Kim}\ \emph {et~al.}(2002)\citenamefont {Kim},
  \citenamefont {Son}, \citenamefont {Bu{\v{z}}ek},\ and\ \citenamefont
  {Knight}}]{Kim:2002a}%
  \BibitemOpen
  \bibfield  {author} {\bibinfo {author} {\bibfnamefont {M.~S.}\ \bibnamefont
  {Kim}}, \bibinfo {author} {\bibfnamefont {W.}~\bibnamefont {Son}}, \bibinfo
  {author} {\bibfnamefont {V.}~\bibnamefont {Bu{\v{z}}ek}}, \ and\ \bibinfo
  {author} {\bibfnamefont {P.~L.}\ \bibnamefont {Knight}},\ }\href {\doibase
  10.1103/PhysRevA.65.032323} {\bibfield  {journal} {\bibinfo  {journal} {Phys.
  Rev. A}\ }\textbf {\bibinfo {volume} {65}},\ \bibinfo {pages} {032323}
  (\bibinfo {year} {2002})}\BibitemShut {NoStop}%
\bibitem [{\citenamefont {Lanzagorta}(2012)}]{Lanzagorta:2012}%
  \BibitemOpen
  \bibfield  {author} {\bibinfo {author} {\bibfnamefont {M.}~\bibnamefont
  {Lanzagorta}},\ }in\ \href {\doibase doi:10.2200/S00384ED1V01Y201110QMC005}
  {\emph {\bibinfo {booktitle} {Quantum Radar}}},\ \bibinfo {series} {Synthesis
  Lectures on Quantum Computing}, Vol.~\bibinfo {volume} {5},\ \bibinfo
  {editor} {edited by\ \bibinfo {editor} {\bibfnamefont {M.}~\bibnamefont
  {Lanzagorta}}\ and\ \bibinfo {editor} {\bibfnamefont {J.}~\bibnamefont
  {Uhlmann}}}\ (\bibinfo  {publisher} {Morgan \& Claypool Publishers},\
  \bibinfo {year} {2012})\BibitemShut {NoStop}%
\end{thebibliography}

\begin{thebibliography}{22}%
\makeatletter
\global\c@NAT@ctr 30\relax
\makeatother
\makeatletter
\providecommand \@ifxundefined [1]{%
 \@ifx{#1\undefined}
}%
\providecommand \@ifnum [1]{%
 \ifnum #1\expandafter \@firstoftwo
 \else \expandafter \@secondoftwo
 \fi
}%
\providecommand \@ifx [1]{%
 \ifx #1\expandafter \@firstoftwo
 \else \expandafter \@secondoftwo
 \fi
}%
\providecommand \natexlab [1]{#1}%
\providecommand \enquote  [1]{``#1''}%
\providecommand \bibnamefont  [1]{#1}%
\providecommand \bibfnamefont [1]{#1}%
\providecommand \citenamefont [1]{#1}%
\providecommand \href@noop [0]{\@secondoftwo}%
\providecommand \href [0]{\begingroup \@sanitize@url \@href}%
\providecommand \@href[1]{\@@startlink{#1}\@@href}%
\providecommand \@@href[1]{\endgroup#1\@@endlink}%
\providecommand \@sanitize@url [0]{\catcode `\\12\catcode `\$12\catcode
  `\&12\catcode `\#12\catcode `\^12\catcode `\_12\catcode `\%12\relax}%
\providecommand \@@startlink[1]{}%
\providecommand \@@endlink[0]{}%
\providecommand \url  [0]{\begingroup\@sanitize@url \@url }%
\providecommand \@url [1]{\endgroup\@href {#1}{\urlprefix }}%
\providecommand \urlprefix  [0]{URL }%
\providecommand \Eprint [0]{\href }%
\providecommand \doibase [0]{http://dx.doi.org/}%
\providecommand \selectlanguage [0]{\@gobble}%
\providecommand \bibinfo  [0]{\@secondoftwo}%
\providecommand \bibfield  [0]{\@secondoftwo}%
\providecommand \translation [1]{[#1]}%
\providecommand \BibitemOpen [0]{}%
\providecommand \bibitemStop [0]{}%
\providecommand \bibitemNoStop [0]{.\EOS\space}%
\providecommand \EOS [0]{\spacefactor3000\relax}%
\providecommand \BibitemShut  [1]{\csname bibitem#1\endcsname}%
\let\auto@bib@innerbib\@empty
%</preamble>
\bibitem [{\citenamefont {Yamamoto}\ \emph {et~al.}(2008)\citenamefont
  {Yamamoto}, \citenamefont {Inomata}, \citenamefont {Watanabe}, \citenamefont
  {Matsuba}, \citenamefont {Miyazaki}, \citenamefont {Oliver}, \citenamefont
  {Nakamura},\ and\ \citenamefont {Tsai}}]{Yamamoto:2008a}%
  \BibitemOpen
  \bibfield  {author} {\bibinfo {author} {\bibfnamefont {T.}~\bibnamefont
  {Yamamoto}}, \bibinfo {author} {\bibfnamefont {K.}~\bibnamefont {Inomata}},
  \bibinfo {author} {\bibfnamefont {M.}~\bibnamefont {Watanabe}}, \bibinfo
  {author} {\bibfnamefont {K.}~\bibnamefont {Matsuba}}, \bibinfo {author}
  {\bibfnamefont {T.}~\bibnamefont {Miyazaki}}, \bibinfo {author}
  {\bibfnamefont {W.~D.}\ \bibnamefont {Oliver}}, \bibinfo {author}
  {\bibfnamefont {Y.}~\bibnamefont {Nakamura}}, \ and\ \bibinfo {author}
  {\bibfnamefont {J.~S.}\ \bibnamefont {Tsai}},\ }\href {\doibase
  10.1063/1.2964182} {\bibfield  {journal} {\bibinfo  {journal} {Appl. Phys.
  Lett.}\ }\textbf {\bibinfo {volume} {93}},\ \bibinfo {pages} {042510}
  (\bibinfo {year} {2008})}\BibitemShut {NoStop}%
\bibitem [{\citenamefont {Collin}(2001)}]{Collin:2001a}%
  \BibitemOpen
  \bibfield  {author} {\bibinfo {author} {\bibfnamefont {R.~E.}\ \bibnamefont
  {Collin}},\ }\href@noop {} {\emph {\bibinfo {title} {Foundations for
  Microwave Engineering}}},\ \bibinfo {edition} {2nd}\ ed.\ (\bibinfo
  {publisher} {Wiley and Sons},\ \bibinfo {address} {New York},\ \bibinfo
  {year} {2001})\BibitemShut {NoStop}%
\bibitem [{\citenamefont {Mariantoni}\ \emph {et~al.}(2010)\citenamefont
  {Mariantoni}, \citenamefont {Menzel}, \citenamefont {Deppe}, \citenamefont
  {{Araque Caballero}}, \citenamefont {Baust}, \citenamefont {Niemczyk},
  \citenamefont {Hoffmann}, \citenamefont {Solano}, \citenamefont {Marx},\ and\
  \citenamefont {Gross}}]{Mariantoni:2010a}%
  \BibitemOpen
  \bibfield  {author} {\bibinfo {author} {\bibfnamefont {M.}~\bibnamefont
  {Mariantoni}}, \bibinfo {author} {\bibfnamefont {E.~P.}\ \bibnamefont
  {Menzel}}, \bibinfo {author} {\bibfnamefont {F.}~\bibnamefont {Deppe}},
  \bibinfo {author} {\bibfnamefont {M.~{\'A}.}\ \bibnamefont {{Araque
  Caballero}}}, \bibinfo {author} {\bibfnamefont {A.}~\bibnamefont {Baust}},
  \bibinfo {author} {\bibfnamefont {T.}~\bibnamefont {Niemczyk}}, \bibinfo
  {author} {\bibfnamefont {E.}~\bibnamefont {Hoffmann}}, \bibinfo {author}
  {\bibfnamefont {E.}~\bibnamefont {Solano}}, \bibinfo {author} {\bibfnamefont
  {A.}~\bibnamefont {Marx}}, \ and\ \bibinfo {author} {\bibfnamefont
  {R.}~\bibnamefont {Gross}},\ }\href {\doibase 10.1103/PhysRevLett.105.133601}
  {\bibfield  {journal} {\bibinfo  {journal} {Phys. Rev. Lett.}\ }\textbf
  {\bibinfo {volume} {105}},\ \bibinfo {pages} {133601} (\bibinfo {year}
  {2010})}\BibitemShut {NoStop}%
\bibitem [{\citenamefont {Hoffmann}\ \emph {et~al.}(2010)\citenamefont
  {Hoffmann}, \citenamefont {Deppe}, \citenamefont {Niemczyk}, \citenamefont
  {Wirth}, \citenamefont {Menzel}, \citenamefont {Wild}, \citenamefont {Huebl},
  \citenamefont {Mariantoni}, \citenamefont {Wei{\ss}l}, \citenamefont
  {Lukashenko}, \citenamefont {Zhuravel}, \citenamefont {Ustinov},
  \citenamefont {Marx},\ and\ \citenamefont {Gross}}]{Hoffmann:2010a}%
  \BibitemOpen
  \bibfield  {author} {\bibinfo {author} {\bibfnamefont {E.}~\bibnamefont
  {Hoffmann}}, \bibinfo {author} {\bibfnamefont {F.}~\bibnamefont {Deppe}},
  \bibinfo {author} {\bibfnamefont {T.}~\bibnamefont {Niemczyk}}, \bibinfo
  {author} {\bibfnamefont {T.}~\bibnamefont {Wirth}}, \bibinfo {author}
  {\bibfnamefont {E.~P.}\ \bibnamefont {Menzel}}, \bibinfo {author}
  {\bibfnamefont {G.}~\bibnamefont {Wild}}, \bibinfo {author} {\bibfnamefont
  {H.}~\bibnamefont {Huebl}}, \bibinfo {author} {\bibfnamefont
  {M.}~\bibnamefont {Mariantoni}}, \bibinfo {author} {\bibfnamefont
  {T.}~\bibnamefont {Wei{\ss}l}}, \bibinfo {author} {\bibfnamefont
  {A.}~\bibnamefont {Lukashenko}}, \bibinfo {author} {\bibfnamefont {A.~P.}\
  \bibnamefont {Zhuravel}}, \bibinfo {author} {\bibfnamefont {A.~V.}\
  \bibnamefont {Ustinov}}, \bibinfo {author} {\bibfnamefont {A.}~\bibnamefont
  {Marx}}, \ and\ \bibinfo {author} {\bibfnamefont {R.}~\bibnamefont {Gross}},\
  }\href {\doibase 10.1063/1.3522650} {\bibfield  {journal} {\bibinfo
  {journal} {Appl. Phys. Lett.}\ }\textbf {\bibinfo {volume} {97}},\ \bibinfo
  {pages} {222508} (\bibinfo {year} {2010})}\BibitemShut {NoStop}%
\bibitem [{\citenamefont {Ou}\ \emph {et~al.}(1992)\citenamefont {Ou},
  \citenamefont {Pereira}, \citenamefont {Kimble},\ and\ \citenamefont
  {Peng}}]{Ou:1992a}%
  \BibitemOpen
  \bibfield  {author} {\bibinfo {author} {\bibfnamefont {Z.~Y.}\ \bibnamefont
  {Ou}}, \bibinfo {author} {\bibfnamefont {S.~F.}\ \bibnamefont {Pereira}},
  \bibinfo {author} {\bibfnamefont {H.~J.}\ \bibnamefont {Kimble}}, \ and\
  \bibinfo {author} {\bibfnamefont {K.~C.}\ \bibnamefont {Peng}},\ }\href
  {\doibase 10.1103/PhysRevLett.68.3663} {\bibfield  {journal} {\bibinfo
  {journal} {Phys. Rev. Lett.}\ }\textbf {\bibinfo {volume} {68}},\ \bibinfo
  {pages} {3663} (\bibinfo {year} {1992})}\BibitemShut {NoStop}%
\bibitem [{\citenamefont {Leonhardt}(1997)}]{Leonhardt:1997a}%
  \BibitemOpen
  \bibfield  {author} {\bibinfo {author} {\bibfnamefont {U.}~\bibnamefont
  {Leonhardt}},\ }\href@noop {} {\emph {\bibinfo {title} {Measuring the Quantum
  State of Light}}}\ (\bibinfo  {publisher} {Cambridge University Press},\
  \bibinfo {address} {Cambridge},\ \bibinfo {year} {1997})\BibitemShut
  {NoStop}%
\bibitem [{\citenamefont {Peropadre}\ \emph {et~al.}(2011)\citenamefont
  {Peropadre}, \citenamefont {Romero}, \citenamefont {Johansson}, \citenamefont
  {Wilson}, \citenamefont {Solano},\ and\ \citenamefont
  {Garc{\'\i}a-Ripoll}}]{Peropadre:2011a}%
  \BibitemOpen
  \bibfield  {author} {\bibinfo {author} {\bibfnamefont {B.}~\bibnamefont
  {Peropadre}}, \bibinfo {author} {\bibfnamefont {G.}~\bibnamefont {Romero}},
  \bibinfo {author} {\bibfnamefont {G.}~\bibnamefont {Johansson}}, \bibinfo
  {author} {\bibfnamefont {C.~M.}\ \bibnamefont {Wilson}}, \bibinfo {author}
  {\bibfnamefont {E.}~\bibnamefont {Solano}}, \ and\ \bibinfo {author}
  {\bibfnamefont {J.~J.}\ \bibnamefont {Garc{\'\i}a-Ripoll}},\ }\href {\doibase
  10.1103/PhysRevA.84.063834} {\bibfield  {journal} {\bibinfo  {journal} {Phys.
  Rev. A}\ }\textbf {\bibinfo {volume} {84}},\ \bibinfo {pages} {063834}
  (\bibinfo {year} {2011})}\BibitemShut {NoStop}%
\bibitem [{\citenamefont {Chen}\ \emph {et~al.}(2011)\citenamefont {Chen},
  \citenamefont {Hover}, \citenamefont {Sendelbach}, \citenamefont {Maurer},
  \citenamefont {Merkel}, \citenamefont {Pritchett}, \citenamefont {Wilhelm},\
  and\ \citenamefont {McDermott}}]{Chen:2012a}%
  \BibitemOpen
  \bibfield  {author} {\bibinfo {author} {\bibfnamefont {Y.-F.}\ \bibnamefont
  {Chen}}, \bibinfo {author} {\bibfnamefont {D.}~\bibnamefont {Hover}},
  \bibinfo {author} {\bibfnamefont {S.}~\bibnamefont {Sendelbach}}, \bibinfo
  {author} {\bibfnamefont {L.}~\bibnamefont {Maurer}}, \bibinfo {author}
  {\bibfnamefont {S.~T.}\ \bibnamefont {Merkel}}, \bibinfo {author}
  {\bibfnamefont {E.~J.}\ \bibnamefont {Pritchett}}, \bibinfo {author}
  {\bibfnamefont {F.~K.}\ \bibnamefont {Wilhelm}}, \ and\ \bibinfo {author}
  {\bibfnamefont {R.}~\bibnamefont {McDermott}},\ }\href {\doibase
  10.1103/PhysRevLett.107.217401} {\bibfield  {journal} {\bibinfo  {journal}
  {Phys. Rev. Lett.}\ }\textbf {\bibinfo {volume} {107}},\ \bibinfo {pages}
  {217401} (\bibinfo {year} {2011})}\BibitemShut {NoStop}%
\bibitem [{\citenamefont {Poudel}\ \emph {et~al.}(2012)\citenamefont {Poudel},
  \citenamefont {McDermott},\ and\ \citenamefont {Vavilov}}]{Poudel:2012a}%
  \BibitemOpen
  \bibfield  {author} {\bibinfo {author} {\bibfnamefont {A.}~\bibnamefont
  {Poudel}}, \bibinfo {author} {\bibfnamefont {R.}~\bibnamefont {McDermott}}, \
  and\ \bibinfo {author} {\bibfnamefont {M.~G.}\ \bibnamefont {Vavilov}},\
  }\href {http://arxiv.org/abs/1201.2990v1} {\bibfield  {journal} {\bibinfo
  {journal} {arXiv:1201.2990v1}\ } (\bibinfo {year} {2012})}\BibitemShut
  {NoStop}%
\bibitem [{\citenamefont {Menzel}\ \emph {et~al.}(2010)\citenamefont {Menzel},
  \citenamefont {Deppe}, \citenamefont {Mariantoni}, \citenamefont {{Araque
  Caballero}}, \citenamefont {Baust}, \citenamefont {Niemczyk}, \citenamefont
  {Hoffmann}, \citenamefont {Marx}, \citenamefont {Solano},\ and\ \citenamefont
  {Gross}}]{Menzel:2010a}%
  \BibitemOpen
  \bibfield  {author} {\bibinfo {author} {\bibfnamefont {E.~P.}\ \bibnamefont
  {Menzel}}, \bibinfo {author} {\bibfnamefont {F.}~\bibnamefont {Deppe}},
  \bibinfo {author} {\bibfnamefont {M.}~\bibnamefont {Mariantoni}}, \bibinfo
  {author} {\bibfnamefont {M.~{\'A}.}\ \bibnamefont {{Araque Caballero}}},
  \bibinfo {author} {\bibfnamefont {A.}~\bibnamefont {Baust}}, \bibinfo
  {author} {\bibfnamefont {T.}~\bibnamefont {Niemczyk}}, \bibinfo {author}
  {\bibfnamefont {E.}~\bibnamefont {Hoffmann}}, \bibinfo {author}
  {\bibfnamefont {A.}~\bibnamefont {Marx}}, \bibinfo {author} {\bibfnamefont
  {E.}~\bibnamefont {Solano}}, \ and\ \bibinfo {author} {\bibfnamefont
  {R.}~\bibnamefont {Gross}},\ }\href {\doibase 10.1103/PhysRevLett.105.100401}
  {\bibfield  {journal} {\bibinfo  {journal} {Phys. Rev. Lett.}\ }\textbf
  {\bibinfo {volume} {105}},\ \bibinfo {pages} {100401} (\bibinfo {year}
  {2010})}\BibitemShut {NoStop}%
\bibitem [{\citenamefont {Bozyigit}\ \emph {et~al.}(2011)\citenamefont
  {Bozyigit}, \citenamefont {Lang}, \citenamefont {Steffen}, \citenamefont
  {Fink}, \citenamefont {Eichler}, \citenamefont {Baur}, \citenamefont
  {Bianchetti}, \citenamefont {Leek}, \citenamefont {Filipp}, \citenamefont
  {da~Silva}, \citenamefont {Blais},\ and\ \citenamefont
  {Wallraff}}]{Bozyigit:2011a}%
  \BibitemOpen
  \bibfield  {author} {\bibinfo {author} {\bibfnamefont {D.}~\bibnamefont
  {Bozyigit}}, \bibinfo {author} {\bibfnamefont {C.}~\bibnamefont {Lang}},
  \bibinfo {author} {\bibfnamefont {L.}~\bibnamefont {Steffen}}, \bibinfo
  {author} {\bibfnamefont {J.~M.}\ \bibnamefont {Fink}}, \bibinfo {author}
  {\bibfnamefont {C.}~\bibnamefont {Eichler}}, \bibinfo {author} {\bibfnamefont
  {M.}~\bibnamefont {Baur}}, \bibinfo {author} {\bibfnamefont {R.}~\bibnamefont
  {Bianchetti}}, \bibinfo {author} {\bibfnamefont {P.~J.}\ \bibnamefont
  {Leek}}, \bibinfo {author} {\bibfnamefont {S.}~\bibnamefont {Filipp}},
  \bibinfo {author} {\bibfnamefont {M.~P.}\ \bibnamefont {da~Silva}}, \bibinfo
  {author} {\bibfnamefont {A.}~\bibnamefont {Blais}}, \ and\ \bibinfo {author}
  {\bibfnamefont {A.}~\bibnamefont {Wallraff}},\ }\href {\doibase
  10.1038/nphys1845} {\bibfield  {journal} {\bibinfo  {journal} {Nature Phys.}\
  }\textbf {\bibinfo {volume} {7}},\ \bibinfo {pages} {154} (\bibinfo {year}
  {2011})}\BibitemShut {NoStop}%
\bibitem [{\citenamefont {Eichler}\ \emph
  {et~al.}(2011{\natexlab{a}})\citenamefont {Eichler}, \citenamefont
  {Bozyigit}, \citenamefont {Lang}, \citenamefont {Steffen}, \citenamefont
  {Fink},\ and\ \citenamefont {Wallraff}}]{Eichler:2011a}%
  \BibitemOpen
  \bibfield  {author} {\bibinfo {author} {\bibfnamefont {C.}~\bibnamefont
  {Eichler}}, \bibinfo {author} {\bibfnamefont {D.}~\bibnamefont {Bozyigit}},
  \bibinfo {author} {\bibfnamefont {C.}~\bibnamefont {Lang}}, \bibinfo {author}
  {\bibfnamefont {L.}~\bibnamefont {Steffen}}, \bibinfo {author} {\bibfnamefont
  {J.}~\bibnamefont {Fink}}, \ and\ \bibinfo {author} {\bibfnamefont
  {A.}~\bibnamefont {Wallraff}},\ }\href {\doibase
  10.1103/PhysRevLett.106.220503} {\bibfield  {journal} {\bibinfo  {journal}
  {Phys. Rev. Lett.}\ }\textbf {\bibinfo {volume} {106}},\ \bibinfo {pages}
  {220503} (\bibinfo {year} {2011}{\natexlab{a}})}\BibitemShut {NoStop}%
\bibitem [{\citenamefont {Eichler}\ \emph
  {et~al.}(2011{\natexlab{b}})\citenamefont {Eichler}, \citenamefont
  {Bozyigit}, \citenamefont {Lang}, \citenamefont {Baur}, \citenamefont
  {Steffen}, \citenamefont {Fink}, \citenamefont {Filipp},\ and\ \citenamefont
  {Wallraff}}]{Eichler:2011b}%
  \BibitemOpen
  \bibfield  {author} {\bibinfo {author} {\bibfnamefont {C.}~\bibnamefont
  {Eichler}}, \bibinfo {author} {\bibfnamefont {D.}~\bibnamefont {Bozyigit}},
  \bibinfo {author} {\bibfnamefont {C.}~\bibnamefont {Lang}}, \bibinfo {author}
  {\bibfnamefont {M.}~\bibnamefont {Baur}}, \bibinfo {author} {\bibfnamefont
  {L.}~\bibnamefont {Steffen}}, \bibinfo {author} {\bibfnamefont {J.~M.}\
  \bibnamefont {Fink}}, \bibinfo {author} {\bibfnamefont {S.}~\bibnamefont
  {Filipp}}, \ and\ \bibinfo {author} {\bibfnamefont {A.}~\bibnamefont
  {Wallraff}},\ }\href {\doibase 10.1103/PhysRevLett.107.113601} {\bibfield
  {journal} {\bibinfo  {journal} {Phys. Rev. Lett.}\ }\textbf {\bibinfo
  {volume} {107}},\ \bibinfo {pages} {113601} (\bibinfo {year}
  {2011}{\natexlab{b}})}\BibitemShut {NoStop}%
\bibitem [{\citenamefont {da~Silva}\ \emph {et~al.}(2010)\citenamefont
  {da~Silva}, \citenamefont {Bozyigit}, \citenamefont {Wallraff},\ and\
  \citenamefont {Blais}}]{DaSilva:2010a}%
  \BibitemOpen
  \bibfield  {author} {\bibinfo {author} {\bibfnamefont {M.~P.}\ \bibnamefont
  {da~Silva}}, \bibinfo {author} {\bibfnamefont {D.}~\bibnamefont {Bozyigit}},
  \bibinfo {author} {\bibfnamefont {A.}~\bibnamefont {Wallraff}}, \ and\
  \bibinfo {author} {\bibfnamefont {A.}~\bibnamefont {Blais}},\ }\href
  {\doibase 10.1103/PhysRevA.82.043804} {\bibfield  {journal} {\bibinfo
  {journal} {Phys. Rev. A}\ }\textbf {\bibinfo {volume} {82}},\ \bibinfo
  {pages} {043804} (\bibinfo {year} {2010})}\BibitemShut {NoStop}%
\bibitem [{\citenamefont {Caves}(1982)}]{Caves:1982a}%
  \BibitemOpen
  \bibfield  {author} {\bibinfo {author} {\bibfnamefont {C.~M.}\ \bibnamefont
  {Caves}},\ }\href {\doibase 10.1103/PhysRevD.26.1817} {\bibfield  {journal}
  {\bibinfo  {journal} {Phys. Rev. D}\ }\textbf {\bibinfo {volume} {26}},\
  \bibinfo {pages} {1817} (\bibinfo {year} {1982})}\BibitemShut {NoStop}%
\bibitem [{\citenamefont {Bu{\v{z}}ek}\ \emph
  {et~al.}(1996{\natexlab{a}})\citenamefont {Bu{\v{z}}ek}, \citenamefont
  {Adam},\ and\ \citenamefont {Drobn{\'y}}}]{Buzek:1996a}%
  \BibitemOpen
  \bibfield  {author} {\bibinfo {author} {\bibfnamefont {V.}~\bibnamefont
  {Bu{\v{z}}ek}}, \bibinfo {author} {\bibfnamefont {G.}~\bibnamefont {Adam}}, \
  and\ \bibinfo {author} {\bibfnamefont {G.}~\bibnamefont {Drobn{\'y}}},\
  }\href {\doibase 10.1103/PhysRevA.54.804} {\bibfield  {journal} {\bibinfo
  {journal} {Phys. Rev. A}\ }\textbf {\bibinfo {volume} {54}},\ \bibinfo
  {pages} {804} (\bibinfo {year} {1996}{\natexlab{a}})}\BibitemShut {NoStop}%
\bibitem [{\citenamefont {Bu{\v{z}}ek}\ \emph
  {et~al.}(1996{\natexlab{b}})\citenamefont {Bu{\v{z}}ek}, \citenamefont
  {Adam},\ and\ \citenamefont {Drobn{\'y}}}]{Buzek:1996b}%
  \BibitemOpen
  \bibfield  {author} {\bibinfo {author} {\bibfnamefont {V.}~\bibnamefont
  {Bu{\v{z}}ek}}, \bibinfo {author} {\bibfnamefont {G.}~\bibnamefont {Adam}}, \
  and\ \bibinfo {author} {\bibfnamefont {G.}~\bibnamefont {Drobn{\'y}}},\
  }\href {\doibase 10.1006/aphy.1996.0003} {\bibfield  {journal} {\bibinfo
  {journal} {Ann. Phys.}\ }\textbf {\bibinfo {volume} {245}},\ \bibinfo {pages}
  {37 } (\bibinfo {year} {1996}{\natexlab{b}})}\BibitemShut {NoStop}%
\bibitem [{\citenamefont {Shchukin}\ and\ \citenamefont
  {Vogel}(2005)}]{Shchukin:2005a}%
  \BibitemOpen
  \bibfield  {author} {\bibinfo {author} {\bibfnamefont {E.}~\bibnamefont
  {Shchukin}}\ and\ \bibinfo {author} {\bibfnamefont {W.}~\bibnamefont
  {Vogel}},\ }\href {\doibase 10.1103/PhysRevLett.95.230502} {\bibfield
  {journal} {\bibinfo  {journal} {Phys. Rev. Lett.}\ }\textbf {\bibinfo
  {volume} {95}},\ \bibinfo {pages} {230502} (\bibinfo {year}
  {2005})}\BibitemShut {NoStop}%
\bibitem [{\citenamefont {Adesso}\ and\ \citenamefont
  {Illuminati}(2005)}]{Adesso:2005a}%
  \BibitemOpen
  \bibfield  {author} {\bibinfo {author} {\bibfnamefont {G.}~\bibnamefont
  {Adesso}}\ and\ \bibinfo {author} {\bibfnamefont {F.}~\bibnamefont
  {Illuminati}},\ }\href {\doibase 10.1103/PhysRevA.72.032334} {\bibfield
  {journal} {\bibinfo  {journal} {Phys. Rev. A}\ }\textbf {\bibinfo {volume}
  {72}},\ \bibinfo {pages} {032334} (\bibinfo {year} {2005})}\BibitemShut
  {NoStop}%
\bibitem [{\citenamefont {Hyllus}\ and\ \citenamefont
  {Eisert}(2006)}]{Hyllus:2006a}%
  \BibitemOpen
  \bibfield  {author} {\bibinfo {author} {\bibfnamefont {P.}~\bibnamefont
  {Hyllus}}\ and\ \bibinfo {author} {\bibfnamefont {J.}~\bibnamefont
  {Eisert}},\ }\href {\doibase 10.1088/1367-2630/8/4/051} {\bibfield  {journal}
  {\bibinfo  {journal} {New J. Phys.}\ }\textbf {\bibinfo {volume} {8}},\
  \bibinfo {pages} {51} (\bibinfo {year} {2006})}\BibitemShut {NoStop}%
\bibitem [{\citenamefont {Schack}\ and\ \citenamefont
  {Schenzle}(1990)}]{Schack:1990a}%
  \BibitemOpen
  \bibfield  {author} {\bibinfo {author} {\bibfnamefont {R.}~\bibnamefont
  {Schack}}\ and\ \bibinfo {author} {\bibfnamefont {A.}~\bibnamefont
  {Schenzle}},\ }\href {\doibase 10.1103/PhysRevA.41.3847} {\bibfield
  {journal} {\bibinfo  {journal} {Phys. Rev. A}\ }\textbf {\bibinfo {volume}
  {41}},\ \bibinfo {pages} {3847} (\bibinfo {year} {1990})}\BibitemShut
  {NoStop}%
\bibitem [{\citenamefont {Jiang}(2010)}]{Jiang:2010a}%
  \BibitemOpen
  \bibfield  {author} {\bibinfo {author} {\bibfnamefont {L.-Z.}\ \bibnamefont
  {Jiang}},\ }\href@noop {} {\bibfield  {journal} {\bibinfo  {journal} {Proc.
  of SPIE}\ }\textbf {\bibinfo {volume} {7846}},\ \bibinfo {pages} {784612}
  (\bibinfo {year} {2010})}\BibitemShut {NoStop}%
\end{thebibliography}
\end{document}